\documentclass[universe,review,submit,pdftex,moreauthors]{mdpi} 

\usepackage[version=4]{mhchem}
\usepackage{graphicx,epsf}
\usepackage{graphicx}	
\usepackage{amsmath}	
\usepackage{amssymb}	
\usepackage{newtxtext,newtxmath}
\usepackage{siunitx}
\usepackage{footnote}

\newcommand{\Msun}{{\rm M}_\odot}
\newcommand{\Rsun}{{\rm R}_\odot}

\newcommand{\kms}{\textrm{km}\,\textrm{s}^{-1}}
\newcommand{\mdot}{{\rm M}$_\odot$\,{\rm yr}$^{-1}$}

\DeclareRobustCommand{\ion}[2]{\relax\ifmmode\ifx\testbx\f@series{\mathbf{#1\,\mathsc{#2}}}\else{\mathrm{#1\,\mathsc{#2}}}\fi\else\textup{#1\,{\mdseries\textsc{#2}}}\fi}

\newcommand{\code}[1]{\texttt{#1}}

\def\cmfgen{{\code{CMFGEN}}}

\firstpage{1} 
\pubvolume{1}
\issuenum{1}
\articlenumber{0}
\pubyear{2025}
\copyrightyear{2025}
\datereceived{ } 
\daterevised{ } 
\dateaccepted{ } 
\datepublished{ } 
\hreflink{https://doi.org/} 

\Title{SN 2023ixf: The Closest Supernova of the Decade}

\TitleCitation{SN 2023ixf: The Closest Supernova of the Decade}

\Author{Wynn Jacobson-Gal\'an$^{1,2}$\orcidA{}}

\AuthorNames{Wynn Jacobson-Gal\'an}

\isAPAStyle{%
       \AuthorCitation{Jacobson-Gal\'an, W.}
         }{%
        \isChicagoStyle{%
        \AuthorCitation{Jacobson-Gal\'an, Wynn}
        }{
        \AuthorCitation{Jacobson-Gal\'an, W.}
        }
}

\address{%
$^{1}$ \quad Department of Astronomy and Astrophysics, California Institute of Technology\\
$^{2}$ \quad NASA Hubble Fellow}

\corres{Correspondence: wynnjg@caltech.edu}

\firstnote{Current address: 1200 East California Blvd,
Pasadena, CA 91125}  

\abstract{Supernova 2023ixf occurred on May 18th 2023 in the nearby galaxy Messier 101 ($D \approx 6.85$~Mpc), making it the closest supernova in the last decade. Following its discovery, astronomers around the world rushed to observe the explosion across the electromagnetic spectrum in order to uncover its early-time properties. Based on multi-wavelength analysis during its first year post-explosion, supernova 2023ixf is a type II supernova that interacted with dense, confined circumstellar material in its local environment -- this material being lost from its red supergiant progenitor in the final years before explosion. In this article, we will review the findings of $>$80 studies already published on this incredible event as well as explore how the synthesis of SN~2023ixf observations across the electromagnetic spectrum can be used to constrain both type II supernova explosion physics in addition to the the uncertain mass-loss histories of red supergiant stars in their final years.}


\begin{document}


\section{Discovery}

\begin{figure*}
\includegraphics[width=\textwidth]{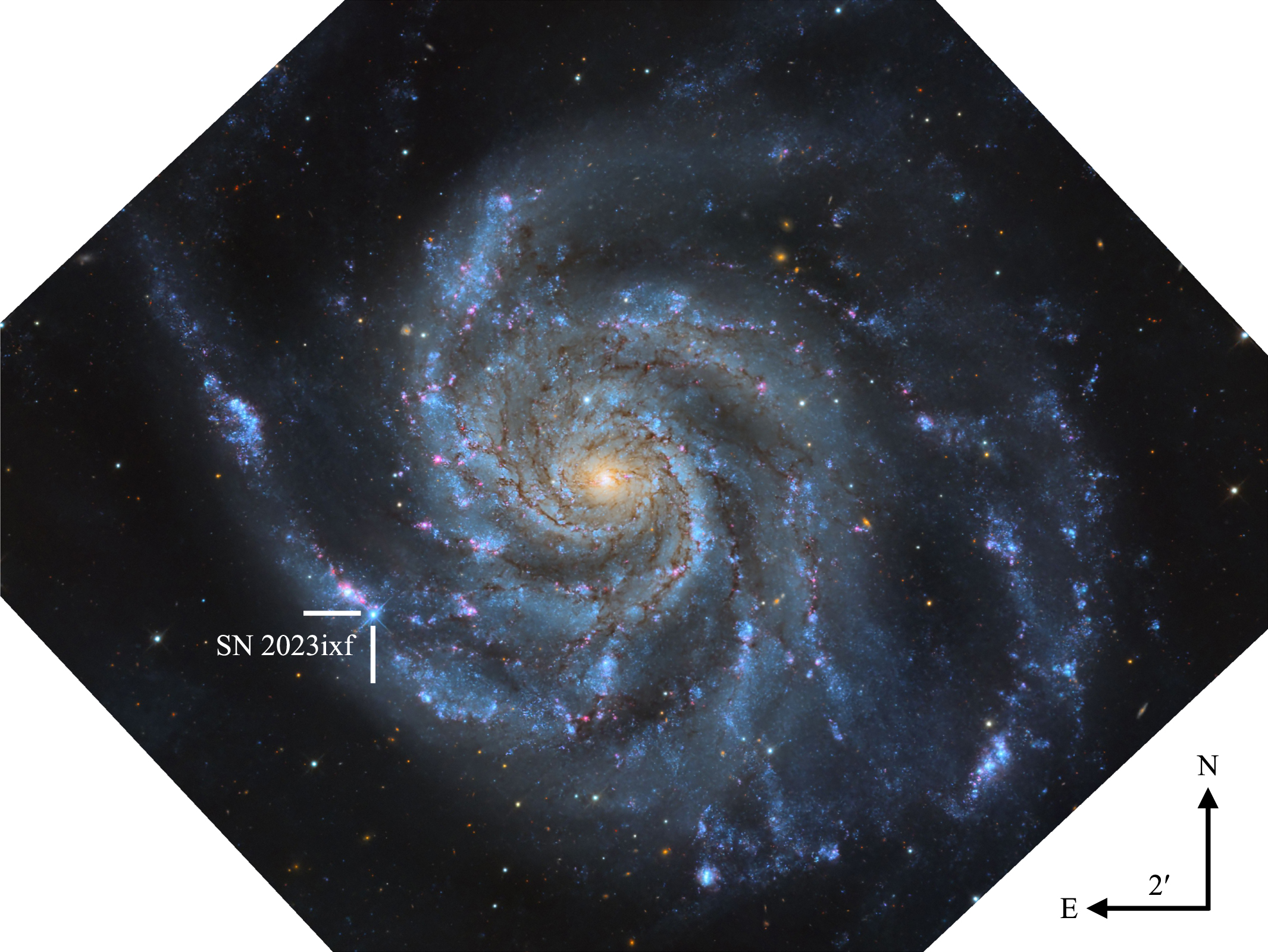}
\caption{SN 2023ixf in its host galaxy Messier 101 (with permission from \cite{Hosseinzadeh23}). Image credit: Travis Deyoe, Mount Lemmon SkyCenter, University of Arizona. \label{fig:image}}
\end{figure*}

The discovery of Supernova (SN) 2023ixf was first reported to the Transient Name Server (TNS) by Koichi Itagaki \cite{Itagaki23} on 2023 May 19 17:27:15 (MJD 60083.73). The discovery magnitude by Itagaki was 14.9 mag in the Clear filter and the SN was located in the southeastern spiral arm of the Pinwheel Galaxy (Messier 101, NGC 5457; Fig. \ref{fig:image}) at a distance of $6.85\pm 0.15$~Mpc \cite{riess22}. Then, on 2023 May 19 22:23:45 (MJD 60083.93), SN~2023ixf was classified as as a type II supernova (SN II) with ``flash ionization lines of H, He, C, and N'' \cite{perley23}. Following discovery and classification, numerous astronomers reported serendipitous detections and non-detections of SN~2023ixf to better estimate the time of first light \cite{Sgro23}. Amongst these observations, the most constraining were reported in \cite{Mao23} with the last deep upper limit being >20.4 mag (< -8.78 mag) on MJD 60082.66 followed by the first detection of $17.1\pm0.1$~mag (-12.1~mag) in $r$-band on MJD 60082.85. Consequently, the time of first light is conservatively placed at MJD $60082.757 \pm 0.097$. However, tighter constraints on the time of first light have been made through model fits to the earliest photometry (e.g., see \cite{Li24}). 

\section{Spectroscopic Properties}

\subsection{CSM-Interaction Phase}

\begin{figure*}
\centering
\includegraphics[width=0.9\textwidth]{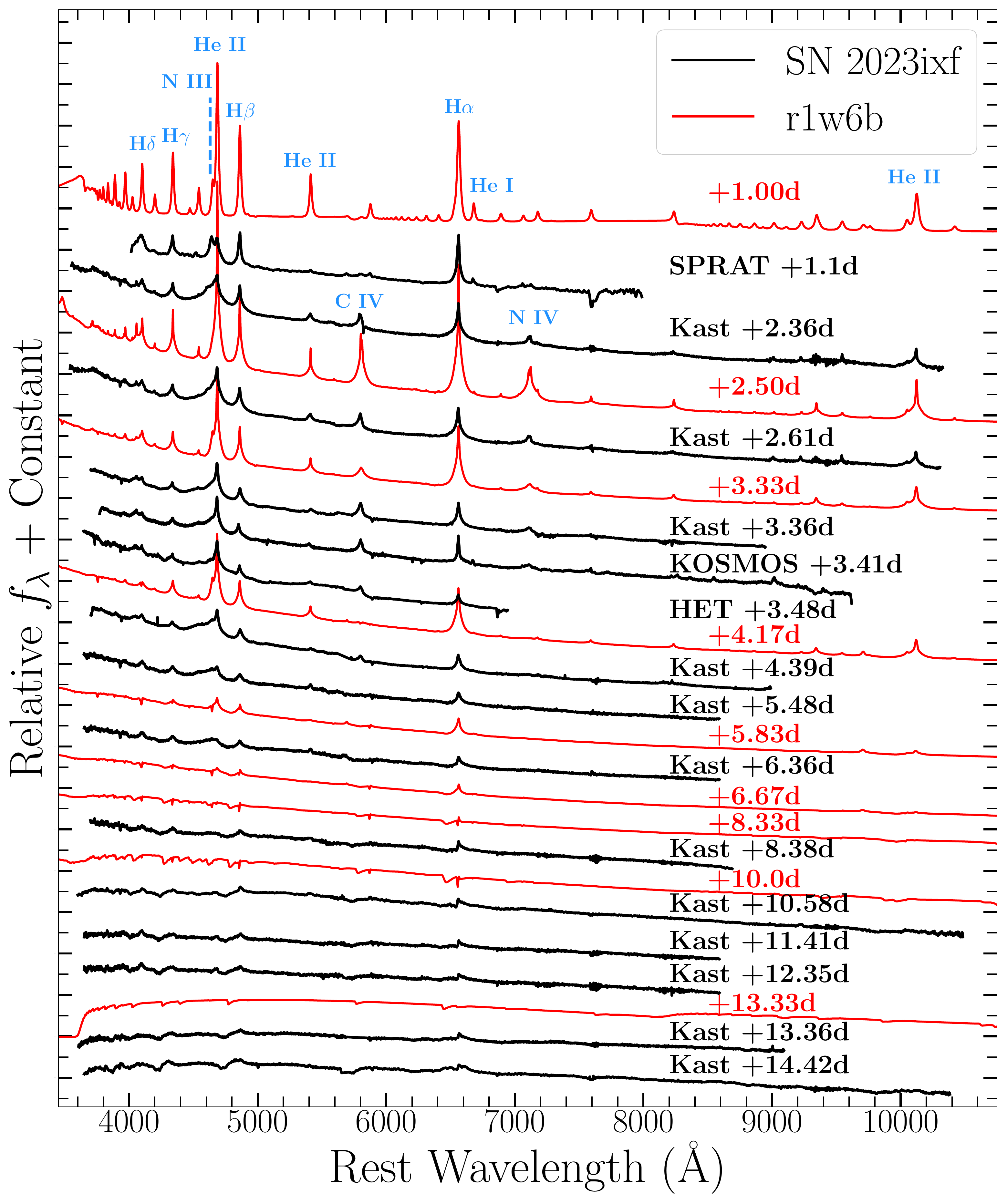}
\caption{Early-time spectral series of SN~2023ixf (black) showing narrow, high-ionization spectral lines from SN ejecta interaction with dense, confined CSM. Best-matched \cmfgen\ spectral model (red) includes a mass loss rate of $10^{-2}$~\mdot, confined to $< 10^{15}$~cm. (Adapted from \cite{wjg23}) \label{fig:spec}}
\end{figure*}

The earliest spectrum of SN~2023ixf was obtained at a phase of $\sim 1.1$~days using the SPRAT on the Liverpool Telescope \cite{perley23}. As shown in Figure \ref{fig:spec}, the early spectra of SN~2023ixf show emission lines of \ion{H}{i}, \ion{He}{i/ii}, \ion{C}{iv}, and \ion{N}{iii/iv/v}, which are superimposed on a hot, blue spectral continuum \cite{wjg23, Bostroem23, Teja23, Zhang23}. Such spectral lines are known as ``flash'' or ``IIn-like'' features and arise from the persistent photo-ionization of dense circumstellar material (CSM) ahead of the cooling forward shock \cite{galyam14, yaron17, dessart17, Khazov16, bruch21, bruch23}. Similar to type IIn SNe, these emission profiles are composed of both a narrow line core and Lorentzian ``wings'' that extend to $\sim1000~\kms$ that result from electron scattering in the ionized, optically-thick CSM \cite{Chugai01, dessart17, Huang18}. However, it should be noted that the velocity of the narrow line core is immediately influenced by radiative acceleration following shock breakout (SBO) and may not trace the exact progenitor wind velocity \cite{Dessart25}. For SN~2023ixf, the earliest high-resolution ($R \approx 70000$) spectroscopy was obtained at $\delta t = 1.51$~days and was used to derive a CSM wind velocity of $\sim 25~\kms$ from the \ion{He}{i} emission line \cite{Dickinson25}. This value is comparable to the narrow line core measurement of SN~II 1998S ($\sim 30 ~\kms$; \cite{shivvers15}), but lower than type IIn and type Ibn SNe ($\sim 100-1000~\kms$; \cite{Smith17}). Continued monitoring with high resolution spectrographs revealed increasing equivalent widths of the narrow line cores that was attributed to either radiative acceleration of the pre-shock gas \cite{Dickinson25, Dessart25} and/or CSM asymmetries \cite{Smith23}. For the latter, the strongest evidence for asymmetric CSM came from spectropolarimetry observations that began at $\delta t = 1.4$~days, the earliest phase that such observations had ever been conducted for a SN to date \cite{Vasylyev23}. The first spectropolarimetric epoch showed high continuum polarization ($p \approx 1\%$), which rapidly decreased over the first week, in addition to a dramatic shift in position angle coincident with fading of the IIn-like features \cite{Vasylyev25}. 

The duration and evolution of IIn-like features in CSM-interacting SNe~II is a direct probe of the CSM structure and shock physics \cite{wjg24a}. Notably, the earliest spectra of SN~2023ixf at $\delta t < 2$~days show electron-scattering broadened emission lines of \ion{He}{i} and \ion{N}{iii}, which fade in strength by $\sim 3$~days as higher ionization species such as \ion{He}{ii}, \ion{N}{iv/v}, and \ion{C}{iv} become increasingly strong \cite{Yamanaka23}. This detection of rising ionization in SN~2023ixf is coincident with increasing blackbody temperature and a red-to-blue color evolution, all of which are attributed to extended shock breakout from dense CSM \cite{Zimmerman24, Li24}. As shown in past studies (e.g., \cite{Dessart23, wjg24a, wjg24b}), the duration of the electron-scattering line profiles can be used to identify the radius at which the pre-shock CSM becomes optically thin to electron-scattering. For SN~2023ixf, this transition occurred at $\delta t \approx 7$~days, which corresponds to a shock radius $\sim 6 \times 10^{14}$~cm for a shock velocity of $\sim 10^4~\kms$ \cite{Bostroem23, wjg23, Teja23, Zimmerman24, zhang24}. At this point, as shown in Figure \ref{fig:spec}, the spectra of SN~2023ixf show only narrow emission in \ion{H}{i} transitions in addition to broad absorption ``troughs,'' indicating that the photosphere had receded into the swept up material present in the fast-moving dense shell (i.e., shocked CSM). Then, after this transition phase, P-Cygni profiles typical of SNe~II emerged from the fastest moving SN ejecta below the dense shell. As multi-wavelength observations in the X-ray and radio revealed, the CSM in SN~2023ixf extends beyond the shock radii when both narrow and electron-scattering broadened emission lines disappear. Estimates on CSM mass, density, and extent as well as progenitor mass-loss rate are discussed in \S\ref{sec:CSM}.

\begin{figure*}
\includegraphics[width=\textwidth]{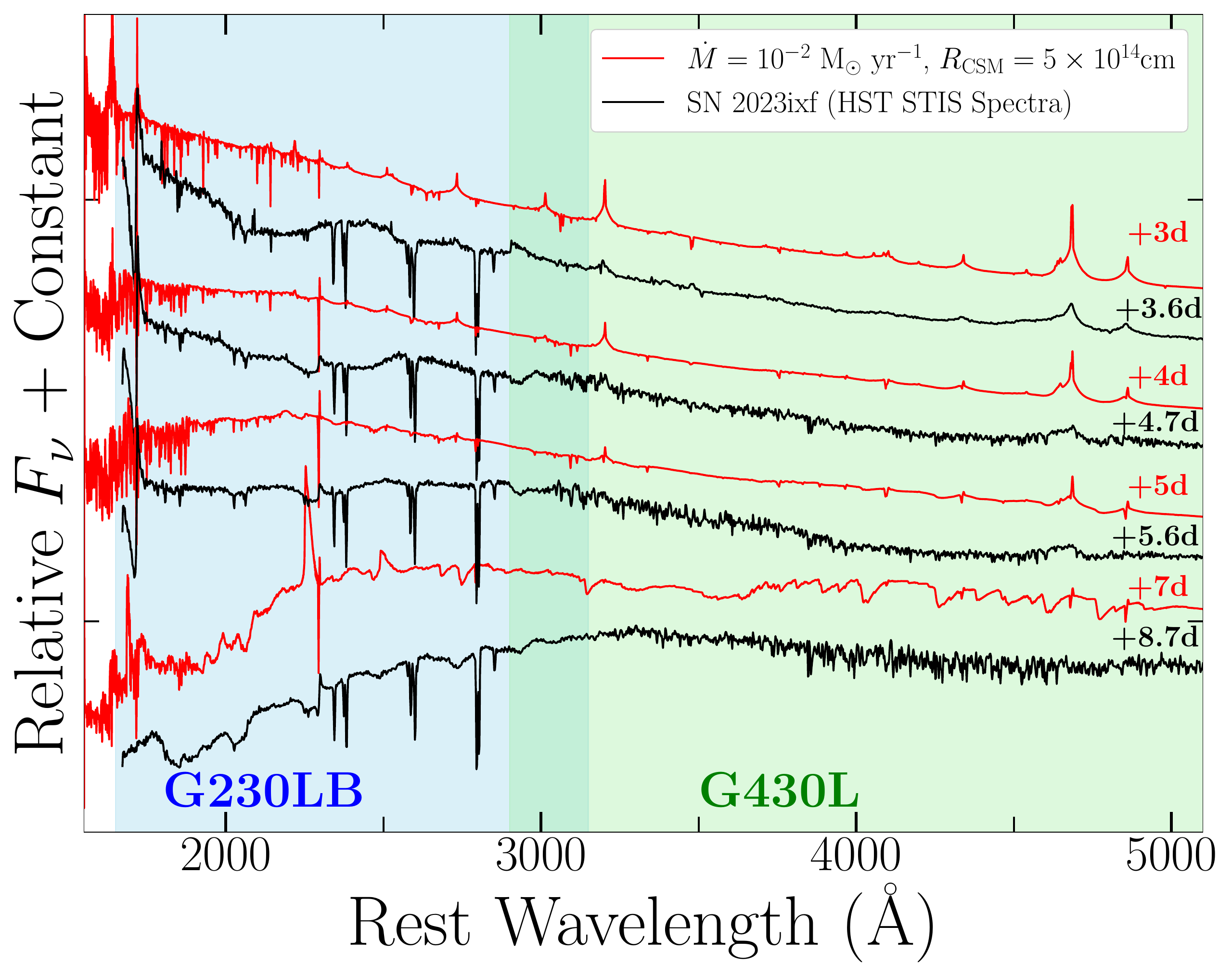}
\caption{Early-time, near-ultraviolet spectroscopy of SN~2023ixf (black) obtained with the {\it HST} STIS CCD in the G230LP (blue region) and G430LP (green region) filters. These observations represented the first detection of narrow \ion{C}{iii} and \ion{N}{iv} emission from CSM-interaction in a young SN~II. \cmfgen\ model spectra from optical spectral matches shown in red. \label{fig:UVspec}}
\end{figure*}

While the majority of the spectroscopic monitoring was in optical wavelengths, SN~2023ixf represented the first successful opportunity to capture the near-ultraviolet (NUV) spectra of a CSM-interacting SN within days of first light. The earliest NUV spectrum was obtained by the Neil Gehrels Swift Observatory at $\delta t = 0.7$~days and showed no obvious spectral features, likely due to the extremely low resolution of the {\it Swift}-UVOT grism \cite{Teja23, Zimmerman24}. Then, as shown in Figure \ref{fig:UVspec}, NUV spectroscopy with the Space Telescope Imaging Spectrograph (STIS) on the {\it Hubble Space Telescope} (HST) was obtained beginning at $\delta t = 3.6$~day \cite{Zimmerman24}. These observations represented the first time that UV spectra of a young SN~II were obtained during the CSM-interaction phase. Intriguingly, the NUV spectrum of SN~2023ixf is relatively featureless -- the strongest predicted IIn-like features are in the far-UV, which is not accessible with the HST/STIS CCD instrument. However, there are detections of narrow \ion{N}{iv} $\lambda1718$ and \ion{C}{iii} $\lambda2297$ emission, the latter of which persists until $\sim 8.7$~days as a combined narrow P-Cygni profile plus broad absorption, similar to H$\alpha$ at the same phase. Additionally, far-UV spectra were obtained as early as +6.9~days using UltraViolet Imaging Telescope (UVIT) on AstroSat \cite{Teja23}. Future UV missions such as the Ultraviolet Explorer (UVEX) will enable routine far- and near-UV spectroscopy of CSM-interacting SNe~II during the first days post-explosion \cite{Kulkarni21}. 

\subsection{Photospheric Phase}

Following the CSM-interaction phase, SN~2023ixf evolved to display Doppler-broadened P-Cygni profiles typical of SNe~II during and after maximum light. As discussed in \cite{Singh24}, the photospheric velocities of SN~2023ixf are overall consistent with samples of SNe~II. Notably, the photospheric phase spectra of SN~2023ixf show some complexities, specifically in the H$\alpha$ profile, such as shallower absorption, high velocity features and boxy redward emission. These observables are proposed to be related to the formation of a cold dense shell (CDS) during the CSM-interaction phase as the forward shock swept up the dense, confined CSM which produced the IIn-like signatures \cite{Chugai07, Dessart22}. Furthermore, the boxy emission on the red side of H$\alpha$ is likely the result of shock power injected into the CDS from continued ejecta interaction with more extended CSM \cite{Zheng25}. The presence of underlying shock power in SN~2023ixf is also validated by the detection of \ion{Mg}{ii} $\lambda 2800$ emission in UV spectra during the photospheric phase \cite{Bostroem24}. 

\subsection{Nebular Phase}

SN~2023ixf began to transition to its nebular phase following the end of the light curve plateau at $\delta t \approx 80$~days. At this point, forbidden emission lines of [\ion{O}{i}] and [\ion{Ca}{ii}] emerged as the SN photosphere receded into the innermost layers of ejecta. Given its brightness, it was possible to track the polarization evolution of SN~2023ixf out to $\sim 120$~days i.e., after the light curve plateau end and at the beginning of the nebular phase. Interestingly, SN~2023ixf displayed increasing polarization at the end of the photospheric phase as the photosphere traces the inner, iron-rich layers of the SN ejecta \cite{Singh24}. This rise in polarization was attributed to an asymmetric distribution of $^{56}$Ni and/or a bipolar explosion \cite{Vasylyev25, Shrestha25}. These findings are broadly consistent with the blueshifted [\ion{O}{i}] emission, which was interpreted as being the result of asymmetric distributions of O-rich ejecta \cite{Fang25}. 

The other notable feature of the late-time spectra in SN~2023ixf is the boxy, asymmetric emission observed in H$\alpha$, which becomes most prominent at $\delta t > 200$~days \cite{Kumar25, Folatelli25, Michel25}. This underlying boxy profile traces emission from the CDS located at $\sim 8000~\kms$ and which is powered by persistent ejecta interaction with distant, intervening CSM at $>10^{15}$~cm. Additionally, the attenuation of redward flux leading to an asymmetric emission in H$\alpha$ is a signature of dust formation in CDS and/or inner SN ejecta. The presence/formation of dust in SN~2023ixf is supported by strong molecular CO emission in the NIR spectra and the presence of SiO emission in the James Webb Space Telescope NIRSpec and MIRI spectra (private communication). 

\section{Photometric Properties}

\begin{figure*}
{\includegraphics[width=0.5\textwidth]{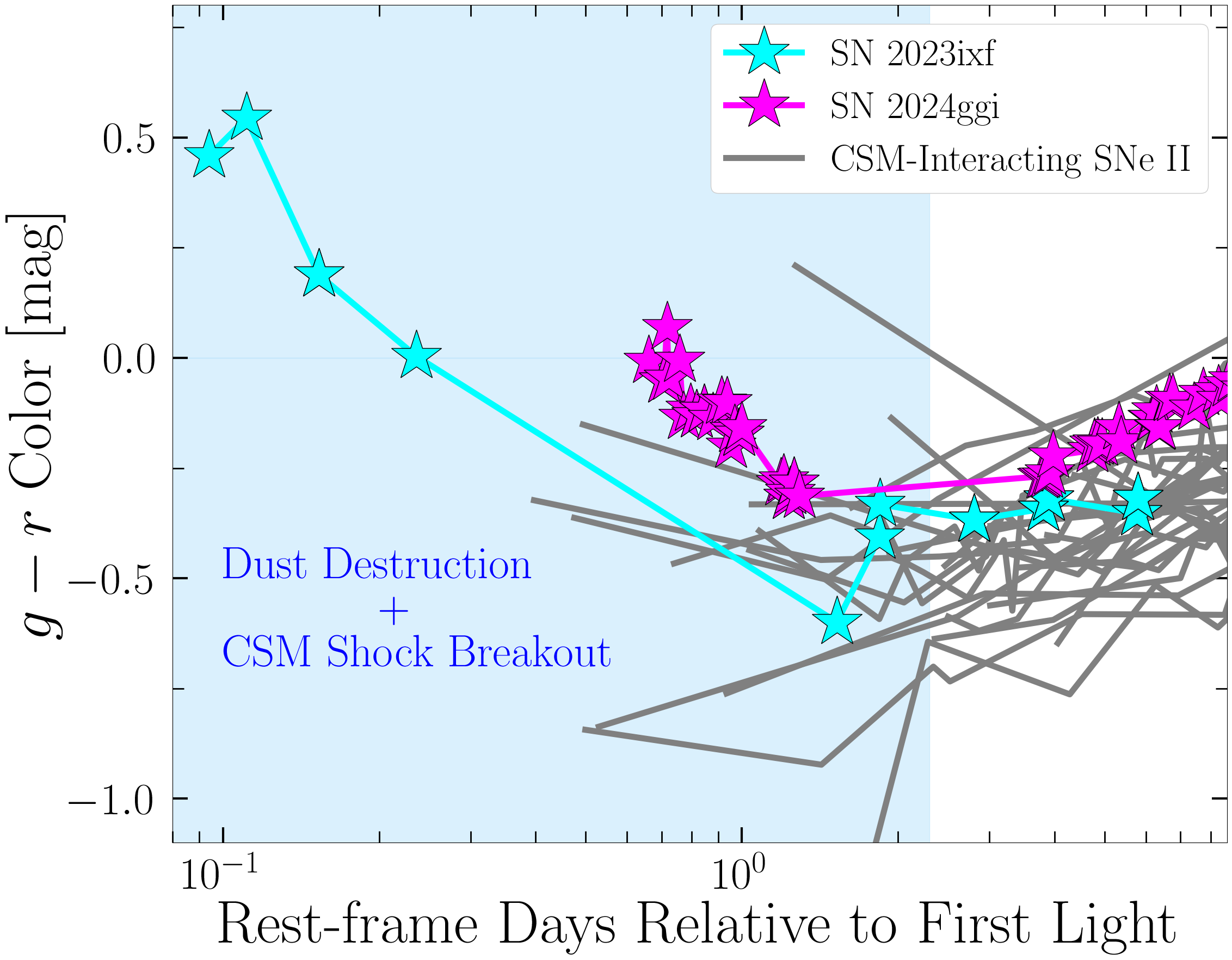}}{\includegraphics[width=0.5\textwidth]{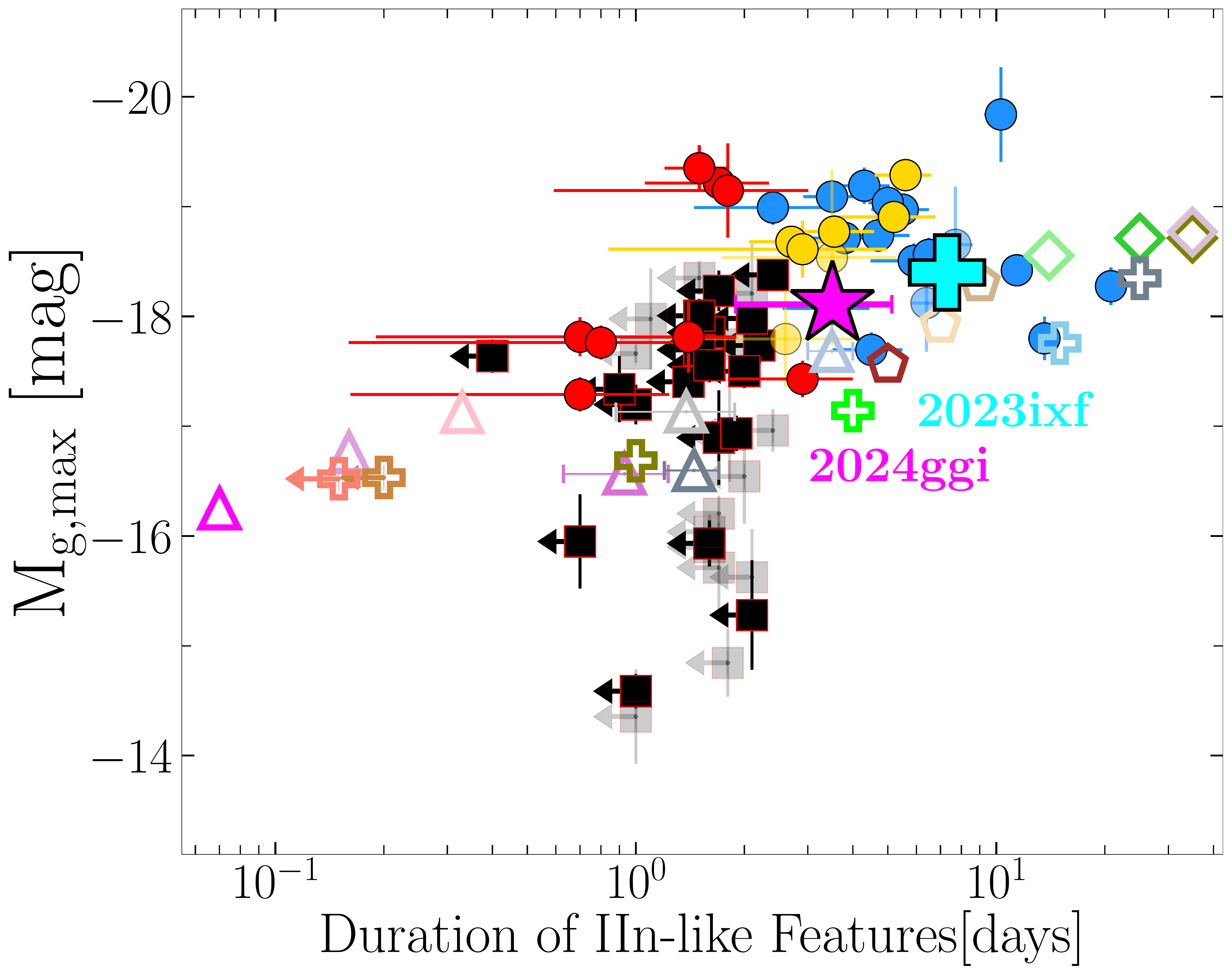}}\\
\caption{{\it Left:} Early-time $g-r$ color evolution of SN~2023ixf (cyan stars) compared to other CSM-interaction SNe~II such as SN~2024ggi (magenta stars; \cite{Shrestha24}) and sample objects (gray lines; \cite{wjg24a}). The dramatic red-to-blue color evolution is proposed to be the product of shock breakout from dense CSM combined with sublimation of the progenitor dust shell \cite{Li24}. {\it Right:} Peak $g$-band absolute magnitude and IIn-like feature durations of SN~2023ixf (cyan plus sign) compared to SN~2024ggi (magenta star; \cite{wjg24b}) and a sample of SNe~II with and without spectroscopic evidence for CSM-interaction at early-times. SN~2023ixf is more luminous than typical SNe~II given its interaction with dense, confined CSM. (Adapted from \cite{wjg24b})   \label{fig:LC}}
\end{figure*} 


Multi-band photometry of SN~2023ixf began $\sim$hours after first light and was able to capture the emergence of the SN shock from within the dense, confined CSM within $\sim 2 R_{\star}$. One of the most compelling features of the SN~2023ixf light curve is the multi-component rise wherein the first $\sim 12$~hours of evolution require a separate power law function fit while the rise to peak brightness at $\delta t > 12$~hours can be described by a typical $F_{\nu} \propto t^2$ model \cite{Hosseinzadeh23}. By utilizing a multi-component light curve model formalism, \cite{Li24} derive a time of first light of MJD $60082.788^{+0.02}_{-0.05}$. Furthermore, during the first $\sim$day, SN~2023ixf displayed a dramatic red-to-blue color evolution (e.g., Fig. \ref{fig:LC}) that appears to be common in CSM-interacting SNe~II with IIn-like features (e.g., \cite{wjg24b, Shrestha24}) and is likely connected to shock breakout from the most confined CSM. This behavior is coincident with a nearly constant blackbody radius and rising blackbody temperature in SN~2023ixf during its first days \cite{Zheng25, Zimmerman24}. Furthermore, \cite{Li24} attribute the reddened colors of SN~2023ixf in its first hours to dust destruction during the CSM shock breakout. Interestingly, this scenario is also consistent with the IR excess observed in SN~2023ixf at $\delta t = 3.6$~days by the serendipitous observations with the Near-Earth Object Wide-field Infrared Survey Explorer (NEOWISE-R) \cite{vandyk24}. 

SN~2023ixf rose to maximum brightness in optical bands within $\sim 5-6$~days and reached an absolute magnitude in $B$-band of $M = -18.5$~mag \cite{wjg23, Singh24, Hinds25}. This enhanced peak luminosity is typical of CSM-interacting SNe~II (e.g., see Fig. \ref{fig:LC}), which can be $1-2$~magnitudes brighter than SNe~II without IIn-like features. Given the linear decline of SN~2023ixf's light curve plateau, it was given the ``type II-L'' SN distinction, which is often physically related to a smaller progenitor star hydrogen envelope mass \cite{Zheng25, Yang24}. SN~2023ixf has a measured pseudo-bolometric light curve plateau duration of $t_p \approx 82$~days, which places it within the sub-class of ``short plateau'' SNe~II \cite{Hsu24, Forde25}. Multiple measurements of the $^{56}$Ni mass were made for SN~2023ixf, with a favored value of $\sim 0.06~\Msun$ estimated from modeling of the post-plateau light curve \cite{Zimmerman24, Singh24, Li25}. Additionally, the light curve ``tail'' of SN~2023ixf declines faster than expected for radioactive decay power with complete $\gamma$-ray trapping and modeling of this evolution recovers a trapping timescale of $\sim 250-300$~days \cite{Singh24, Li25}. Presently, at $\delta t > 600$~days, the optical light curve of SN~2023ixf has flattened, deviating from the expected decline rate of radioactive decay power, as shock power becomes dominant. This phenomenon was also observed at earlier phases in UV and X-ray wavelengths (Jacobson-Gal\'an et al., in prep.). 


\section{X-ray Observations}

\begin{figure*}
{\includegraphics[width=0.55\textwidth]{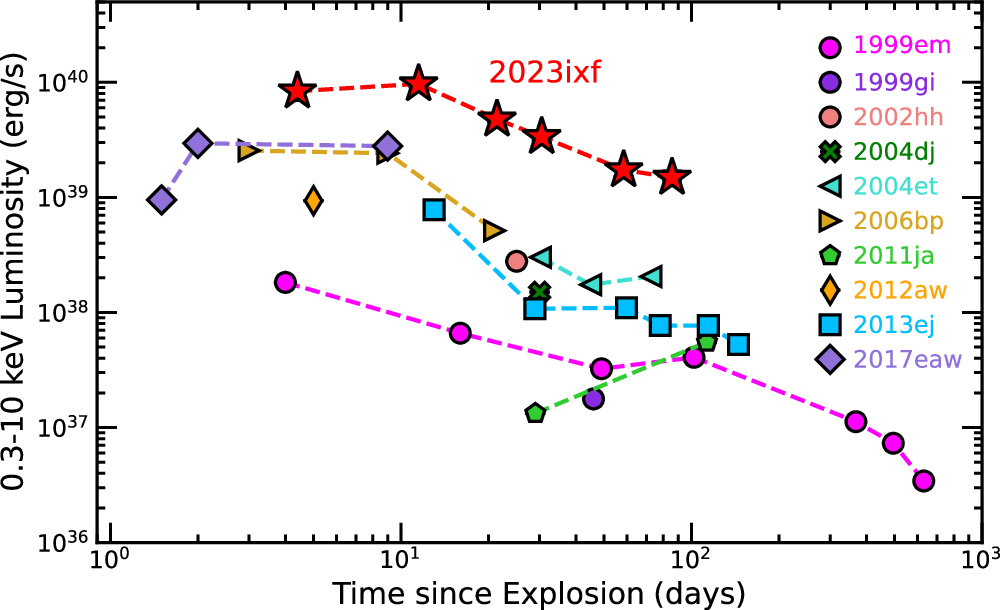}}{\includegraphics[width=0.45\textwidth]{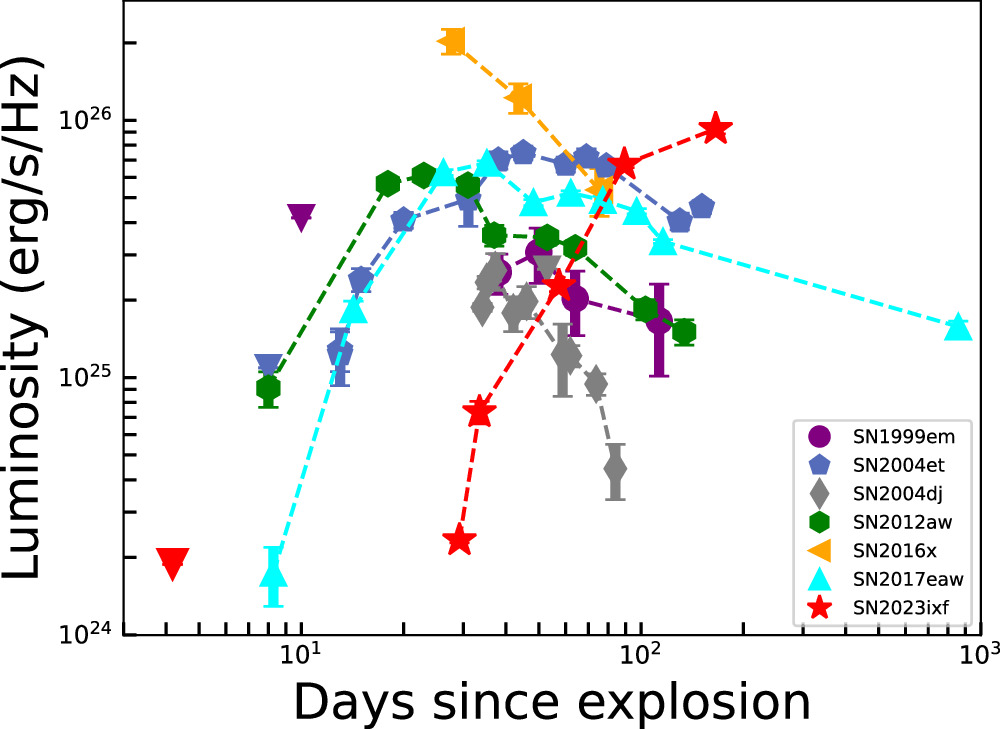}}\\
\caption{{\it Left:} Unabsorbed X-ray luminosity of SN~2023ixf (red stars) compared to other X-ray detected SNe~II. {\it Right:} Multi-frequency radio luminosity of SN~2023ixf compared to other radio SNe~II. (With permission from \cite{Nayana25}) \label{fig:XrayRadio}}
\end{figure*}

SN~2023ixf was first observed at X-ray wavelengths by {\it Swift}-XRT (0.3-10~keV) beginning $\sim 1$~days after first light. Additionally, high signal-to-noise X-ray spectra were obtained using {\it NuSTAR} (3-79~keV) at $\delta t = 4.4 - 58.4$~days, {\it Chandra} (0.5-8~keV) at $\delta t = 11.5 - 86.7$~days, and {\it XMM-Newton} (0.3-10~keV) at $\delta t = 9.0 - 58.2$~days. The exceptional coverage of SN~2023ixf in both soft and hard X-ray bands allowed for robust constraints on the temperature of the X-ray emitting plasma -- this being the first time that such observations were possible for a SNe~II with IIn-like features \cite{Grefenstette23, Chandra24, Panjkov24}. The X-ray emission observed in SN~2023ixf during the first $\sim 90$~days was dominated by the forward shock and is best modeled as an absorbed bremsstrahlung spectrum \cite{Nayana25, Chandra24}. As shown in Figure \ref{fig:XrayRadio}, SN~2023ixf rose to a peak X-ray luminosity of $\sim 10^{40}$~erg~s$^{-1}$ in $\sim1$~week, and represents one of the most luminous SN~II-P/L observed to date. The detection of prominent neutral Fe K$\alpha$ emission combined with large intrinsic neutral hydrogen column densities from X-ray spectral modeling both confirm absorption of X-rays by dense, confined CSM within the first days to a week post-explosion \cite{Grefenstette23, Nayana25}. The rise to maximum X-ray brightness is attributed to decreasing photoelectric absorption as the CSM densities ahead of the forward shock also decrease in time. Currently, SN~2023ixf continues to be detected with X-ray telescopes and the evolution of the X-ray spectrum may indicate the emergence of the radiative reverse shock (private communication).

\section{Radio Observations}

A high cadence, multi-frequency campaign was also carried out in the radio with observations beginning as early as $\delta t = 2.6$~days with the Submillimeter Array (SMA). However, radio emission was not detected until $\delta t = 29.2$~days using the Karl G. Jansky Very Large Array (VLA) \cite{Berger23, Matthews23}. Additional observations were made with the Giant Metrewave Radio Telescope (GMRT), LOw Frequency ARray (LOFAR), the Northern Extended Millimetre Array (NOEMA), the Japanese and Korean VLBI Networks, and the European VLBI Network (EVN) \cite{Nayana25, Timmerman24, Iwata25, Lee24}. The radio emission in SN~2023ixf is attributed to non-thermal synchrotron radiation from the forward shock that is suppressed by free-free absorption, which decreases in time as the density of the CSM also decreases \cite{Nayana25}. 

\section{Neutrinos, $\gamma$-rays and Gravitational Waves}

SN~2023ixf allowed for some of the best constraints on neutrino, $\gamma$-ray and gravitational wave (GW) emission from a SN to date, despite the non-detection of all multi-messenger signals. Specifically, the CSM-interaction in SN~2023ixf provided a framework for the production of high-energy cosmic rays (CRs), $\gamma$-ray, and neutrinos. Overall, non-detection limits from  Fermi-LAT and IceCube proved consistent with the expectations for neutrino and $\gamma$-ray generation using the CSM and explosion parameters inferred from X-ray through radio observations \cite{Marti24, Guetta23, Sarmah24, Kimura25, Ravensburg24, Cosentino25}. Lastly, LIGO-Virgo-KAGRA did not detect any GWs coincident with SN~2023ixf but those observations were able to place the most stringent constraints to date on core-collapse SN GW energy/luminosity and proto-NS ellipticity \cite{Abac25}.

\section{Progenitor System}\label{sec:progenitor}

\subsection{Pre-Explosion Imaging}

\begin{figure*}
{\includegraphics[width=0.5\textwidth]{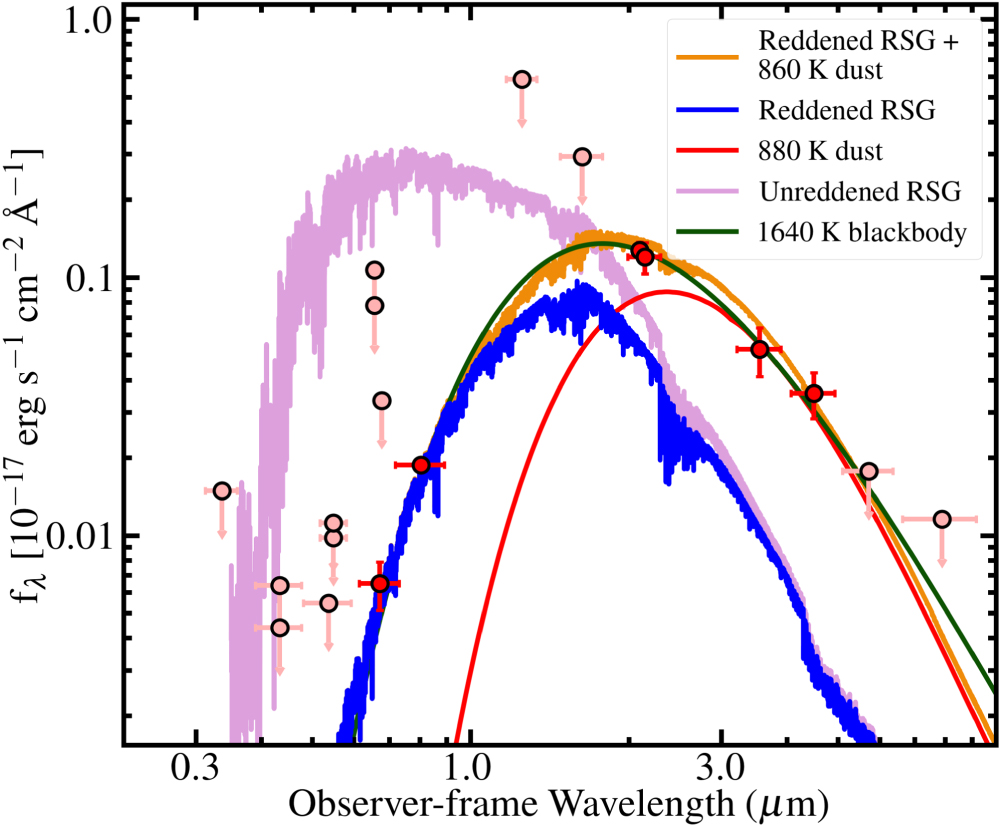}}{\includegraphics[width=0.5\textwidth]{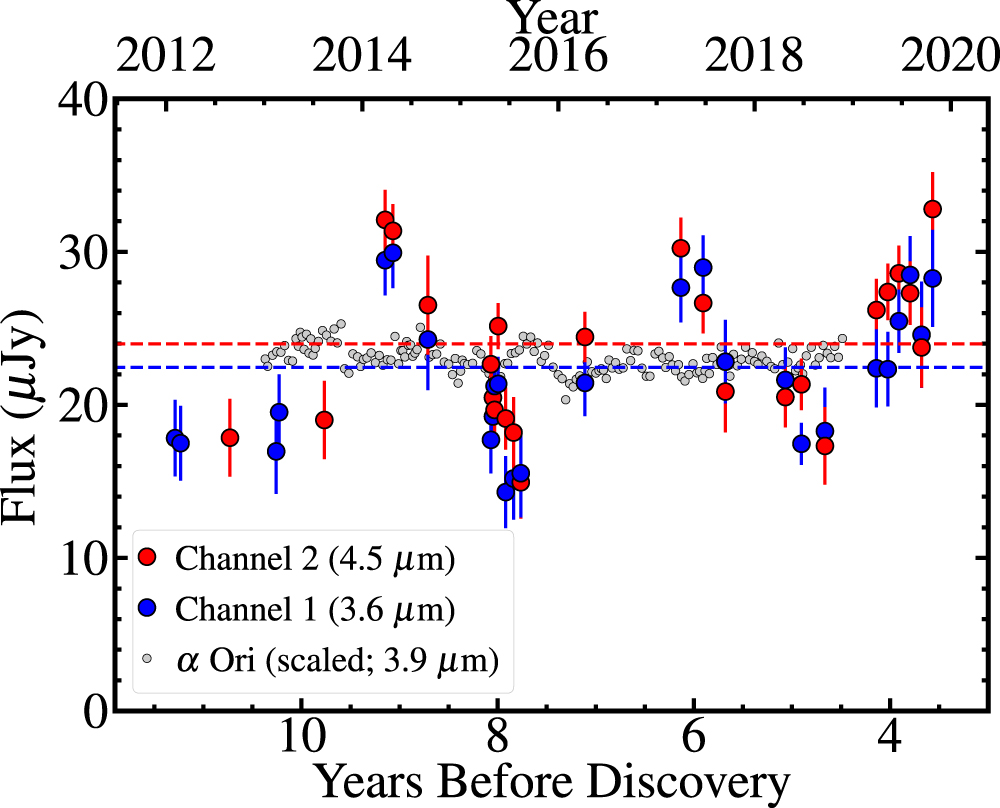}}\\
\caption{ {\it Left:} Progenitor star spectral energy distribution constructed from pre-explosion images of SN~2023ixf. The progenitor star was confirmed to be a red supergiant that was enshrouded in a thick dust shell. {\it Right:} Infrared pre-explosion photometry showing dramatic variability of the progenitor star before explosion. (With permission from \cite{Kilpatrick23})  \label{fig:progenitor}}
\end{figure*}

Given its occurance in M101, SN~2023ixf had a significant amount of associated pre-explosion imaging with both {\it HST} and {\it Spitzer}, which extended more than two decades before first light. As shown in Figure \ref{fig:progenitor}, a progenitor star was clearly detected in some optical {\it HST} filters and Channels 1 \& 2 of {\it Spitzer}. Additional $J$- and $K$-band detections of the progenitor star were made through the Near-Infrared Imager (NIRI) on the Gemini-North Telescope and the MMT Observatory telescope. Based on the pre-explosion spectral energy distribution (SED), it was clear that the progenitor star was a red supergiant (RSG) that was enshrouded in a dust shell as evidenced by the highly reddened SED (e.g., Fig. \ref{fig:progenitor}). However, estimates on Zero Age Main Sequence (ZAMS) based on the progenitor luminosity and temperature are wide-ranging: $11\pm 1~\Msun$ \cite{Kilpatrick23}, $17\pm 4~\Msun$ \cite{Jencson23}, $17\pm 2~\Msun$ \cite{Niu23}, $20\pm 4~\Msun$ \cite{Soraisam23}, $13\pm 1~\Msun$ \cite{VanDyk23}, $18\pm 1~\Msun$ \cite{Qin23}, $\sim 8-10~\Msun$ \cite{Pledger23}, $12\pm 2~\Msun$ \cite{Xiang24}, and $17\pm 3~\Msun$ \cite{Ransome24}. Analysis of the progenitor SED showed no detection of a binary companion but given the non-detection limits in optical filters, only secondary stars with masses $>6.4~\Msun$ could be ruled out. Intriguingly, NIR imaging with {\it Spitzer} showed significant variability with an estimated pulsation period of $\sim 1000$~days \cite{Kilpatrick23, Jencson23, Soraisam23}. Given the strong CSM-interaction during the early evolution of SN~2023ixf, comprehensive searches were performed to look for precursor events, which have been detected prior to some CSM-interacting SNe (e.g., \cite{wjg22, Strotjohann21}). However, no optical ground based survey detected any precursor emission for SN~2023ixf, which effectively ruled out any outbursts or eruptions from the RSG progenitor star in the final years before explosion \cite{Panjkov24, Neustadt24, Ransome24, Dong23, Rest25, Flinner23}. 

\subsection{Light Curve and Spectral Modeling}

Progenitor mass estimates were also made through modeling of the light curve and nebular spectra of SN~2023ixf. Using hydrodynamical modeling of the bolometric light curve during and after the plateau, \cite{Bersten24} favor a $M_{\rm ZAMS} = 12~\Msun$ progenitor with radius of $720~\Rsun$ that explodes with kinetic energy of 1.2~B. Similarly, \cite{Moriya24} and \cite{Singh24} prefer the explosion of a $M_{\rm ZAMS} = 10~\Msun$ progenitor. Alternatively, \cite{Hsu24} and \cite{Forde25} favor a larger progenitor ($M_{\rm ZAMS} = 17~\Msun$) with an inflated radius that had lost significant mass throughout its lifetime, leading to a depleted H envelope mass. In addition to light curve modeling, comparison to spectral models from \cite{Jerkstrand14} and \cite{Dessart21} at nebular phases was also used to estimate the ZAMS mass of the SN~2023ixf RSG progenitor star. However, model matching to nebular spectra also yielded a variety of preferred progenitor masses: $<12~\Msun$ \cite{Kumar25}, $12-15~\Msun$ \cite{Ferrari24}, $<15~\Msun$ \cite{Michel25}, $15.2-16.3~\Msun$ \cite{Fang25}, $15-19~\Msun$ \cite{Li25}, $10-15~\Msun$ \cite{Folatelli25} and $12.5-15~\Msun$ \cite{wjg25a}. 

\subsection{CSM Structure and Mass-Loss Mechanisms}\label{sec:CSM}

Multi-wavelength observations of SN~2023ixf throughout its evolution enabled separate estimates on the circumstellar density profile and the mass loss rate of the RSG progenitor in the final years to decades prior to explosion. The presence of transient, high-ionization emission lines with electron-scattering wings in the early-time SN~2023ixf spectra necessitates CSM optical depths of $\tau \approx 3-10$, which corresponds to a CSM density of $\sim 10^{-12}$~g~cm$^{-3}$ at $10^{14}$~cm. Furthermore, applying a IIn-like feature timescale of $\sim 7$~days and a shock radius of $10^4~\kms$ requires that the transition region from optically thick to thin pre-shock CSM occurs at $\sim 6\times 10^{14}$~cm; this corresponds to a lookback time of $\sim 8$~years pre-explosion for a wind velocity of $25~\kms$. As shown in Figure \ref{fig:spec}, the early-time spectral evolution is well matched by \cmfgen\ model spectra generated for RSG explosion interacting with a enhanced mass loss rate of $10^{-2}~$\mdot ($v_w = 50~\kms$), confined to a radius of $<10^{15}$~cm, with a total CSM mass in the range of $0.04- 0.07~\Msun$. Notably, the spectral models can match all optical emission line species present in SN~2023ixf without CNO enrichment in the CSM. Additionally, \cite{Vasylyev25} model the spectropolarimetry observations of SN~2023ixf using 2D polarized \cmfgen\ models to find that the early-time polarization can be explained by confined CSM with a pole-to-equator density contrast of $\sim$3. 

\begin{figure*}
\includegraphics[width=\textwidth]{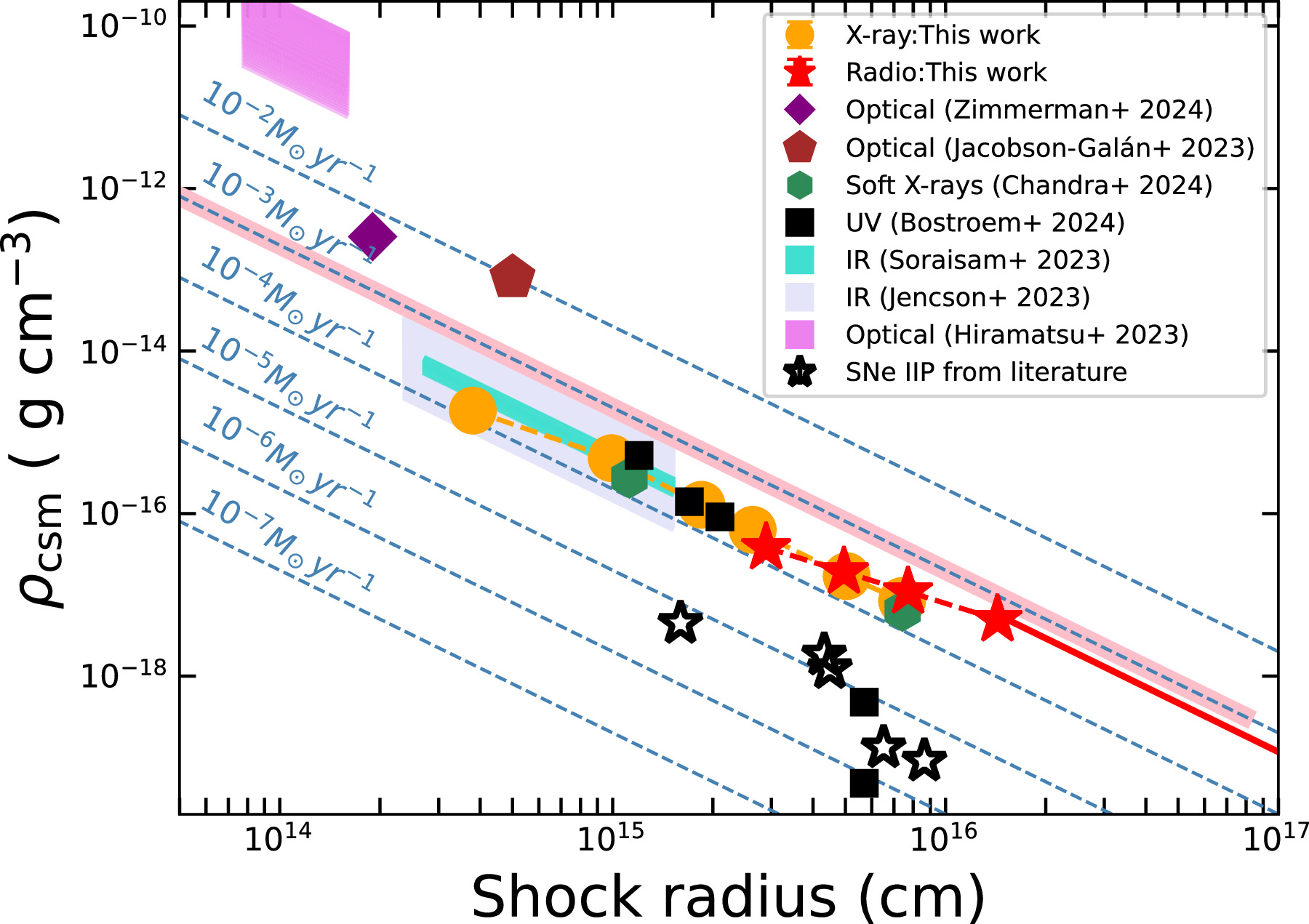}
\caption{ Circumstellar density profile of the SN~2023ixf progenitor star constructed from mass loss rate estimates across the electromagnetic spectrum. Overall, RSG progenitor star was enshrouded in high density CSM at $< 5 \times 10^{14}$~cm, which was formed in the last $\sim 3-6$~years before explosion. At larger radii and pre-SN lookback times, the SN~2023ixf appeared to have an enhanced wind with a continuous mass loss rate of $\sim 10^{-4}$~\mdot. (With permission from \cite{Nayana25}) \label{fig:rho}}
\end{figure*}   

Similar to spectral models, light curve modeling of SN~2023ixf also required confined, high-density CSM to explain the fast rise to a luminous peak brightness. For example, \cite{Martinez24} use hydrodynamical modeling of the bolometric light curve to derive a best-fit mass-loss rate of $3\times 10^{-3}$~\mdot, confined to $< 8\times 10^{14}$~cm, and with a wind acceleration parameter of $\beta = 5$. Modeling of the multi-band light curve by \cite{Moriya24} and \cite{Singh24} suggests a multi-component density profile with higher mass loss $\dot M = 10^{-2}$~\mdot\ within $R< 5\times10^{14}$~cm and lower mass loss of $10^{-4}$~\mdot\ at $R> 5\times10^{14}$~cm. A similarly large mass loss rate of $5\times 10^{-2}$~\mdot\ within $R_{\rm CSM} \approx 5 \times 10^{14}$~cm is found from a hydrid shock cooling plus CSM-interaction analytic model employed by \cite{Li24}. A similar combined shock cooling and CSM-interaction model was employed by \cite{Hu25}, who used a Monte Carlo method to simulate radiative diffusion in order to match the earliest photometry of SN~2023ixf with $\dot M = 10^{-2}$~\mdot\ and $R_{\rm CSM} = 10^{15}$~cm. Furthermore, model light curves associated with the best-matched \cmfgen\ models discussed above can reproduce the light curve peak across filters but likely require additional CSM mass directly above the stellar surface to reproduce the fast rise time observed in SN~2023ixf. Additionally, some light curve models require a significant amount of CSM e.g., \cite{Hiramatsu23} finds $\dot M = 0.1 - 1$~\mdot\ for a continuous mass loss scenario and $M_{\rm CSM} = 0.3-1~\Msun$ for an eruption scenario. Similarly, radiative transfer light curve models by \cite{Kozyreva25} require $0.5-0.9~\Msun$ of confined CSM to match the SN~2023ixf light curve. However, as shown by studies such as \cite{Dessart23}, IIn-like spectral features cannot form from SN ejecta interaction with such a large CSM mass within a confined radius.

Beyond UV/optical observations, modeling of the multi-epoch X-ray spectra was used to estimate a progenitor mass loss rate of $\sim 3 \times 10^{-4}$~\mdot\ beginning at shock radii of $\sim 3\times 10^{14}$~cm \cite{Chandra24, Nayana25, Panjkov24}. Furthermore, radio SED fitting confirmed a similar mass loss rate that remained consistent with a wind-like density profile out to radii $>10^{16}$~cm \cite{Nayana25}. Overall, both X-ray and multi-band radio observations modeled by \cite{Nayana25} are in agreement with the mass loss rates inferred from modeling of the progenitor star SED \cite{Kilpatrick23, Soraisam23, Jencson23, Qin23}. Intriguingly, the earliest high frequency radio observations presented by \cite{Berger23} are inconsistent with the CSM densities inferred from the earliest X-ray detections. Similarly, the presence of IIn-like features in optical spectroscopy necessitates a larger CSM density at the same epoch than is inferred from the X-rays. This inconsistency is potentially reconciled through CSM asymmetries such as a clumpy progenitor wind. The complete CSM density profile constructed from all multi-wavelength observations is shown in Figure \ref{fig:rho} and all CSM properties presented in the literature are summarized in Table \ref{tab1}.

\begin{table*}[t!] 
\caption{Estimates from the literature for SN~2023ixf progenitor mass, mass-loss rate, CSM radius and CSM velocity. \label{tab1}}
\begin{adjustwidth}{-\extralength}{0cm}
\begin{tabularx}{\fulllength}{CCCCCC}
\toprule
\textbf{Ref.} & Method & \textbf{$M_{\rm ZAMS}$} & \textbf{$\dot M$} & \textbf{$R_{\rm CSM}$} & \textbf{$v_{\rm CSM}$} \\
 &  & ($\Msun$) & (\mdot) & (cm) & ($\kms$) \\
\midrule
\cite{Kilpatrick23} & Pre-SN imaging & $11\pm 1$ & $(1.3 \pm 0.1) \times 10^{-6}$ & $(5.8 \pm 0.6) \times 10^{14}$ & 50\\
\cite{Jencson23} & Pre-SN imaging & $17\pm 4$ & $(0.3 
 - 3) \times 10^{-4}$ & $> 4 \times 10^{14}$ & 10\\
 \cite{Niu23} & Pre-SN imaging & $17\pm 2$ & $2 \times 10^{-4}$ & $> 2 \times 10^{15}$ & 115\\
\cite{Jencson23} & Pre-SN imaging & $17\pm 4$ & $(0.3 
 - 3) \times 10^{-4}$ & $> 4 \times 10^{14}$ & 10\\
\cite{Soraisam23} & Pre-SN imaging & $20\pm 4$ & $ (2-4) \times 10^{-4}$ & -- & --\\
\cite{VanDyk23} & Pre-SN imaging & $13\pm 1$ & -- & -- & --\\
\cite{Qin23} & Pre-SN imaging & $18\pm 1$ & $(3.3 \pm 0.26) \times 10^{-4}$ & $> 10^{15}$ & 50\\
\cite{Pledger23} & Pre-SN imaging & $8-12$ & -- & -- & --\\
\cite{Xiang24} & Pre-SN imaging & $12\pm 2$ & $(6-9) \times 10^{-6}$ & $(1.7 - 8.1) \times 10^{15}$ & 70\\
\cite{Ransome24} & Pre-SN imaging & $17\pm 3$ & -- & -- & --\\
\cite{Neustadt24} & Pre-SN imaging & $9-14$ & $10^{-5}$ & $5\times 10^{14}$ & 10\\
\cite{Bersten24} & LC model & $12$ & -- & -- & --\\
\cite{Martinez24} & LC model & $12$ & $3\times 10^{-3}$ & $8\times 10^{14}$ & 115\\
\cite{Moriya24} & LC model & $10$ & $10^{-3} - 10^{-2}$ & $(0.6-1)\times 10^{15}$ & 10\\
\cite{Singh24} & LC model & $10$ & $10^{-2}$ & $< 5\times 10^{14}$ & 10\\
\cite{Singh24} & LC model & $10$ & $10^{-4}$ & $> 5\times 10^{14}$ & 10\\
\cite{Hsu24} & LC model & $>17$ & $10^{-1} - 10^{0}$ & $(0.4 - 1)\times 10^{14}$ & 115\\
\cite{Hiramatsu23} & LC model & -- & $10^{-1} - 10^{0}$ & $(3 - 7)\times 10^{15}$ & 115\\
\cite{Teja23} & LC model & -- & $10^{-3}$ & $7\times 10^{14}$ & 10\\
\cite{Forde25} & LC model & $>17$ & -- & -- & --\\
\cite{Li24} & LC model & -- & $5\times 10^{-2}$ & $5 \times 10^{14}$ & 100\\
\cite{Hu25} & LC model & -- & $10^{-2}$ & $10^{15}$ & 75\\
\cite{Zimmerman24} & LC Model & -- & $10^{-1.96}$ & $2 \times 10^{14}$ & 30\\
\cite{wjg23} & Spectra/LC Model & -- & $10^{-2}$ & $(0.5-1) \times 10^{15}$ & 50\\
\cite{Bostroem23} & Spectra Model & -- & $10^{-3} - 10^{-2}$ & $5 \times 10^{14}$ & 50\\
\cite{Bostroem24} & Spectra Model & -- & $10^{-3} - 10^{-5}$ & $(1-6) \times 10^{15}$ & 50\\
\cite{Zhang23} & Spectra Model & -- & $6 \times 10^{-4}$ & $7 \times 10^{14}$ & 55\\
\cite{Vasylyev25} & Spectra Model & -- & $10^{-2}$ & $8 \times 10^{14}$ & 50\\
\cite{Dickinson25} & Spectra Model & -- & $10^{-2}$ & $8 \times 10^{14}$ & 25\\
\cite{Grefenstette23} & X-ray Model & -- & $3 \times 10^{-3}$ & $<10^{15}$ & 50\\
\cite{Chandra24} & X-ray Model & -- & $5 \times 10^{-4}$ & $(0.6 - 4) \times 10^{15}$ & 115\\
\cite{Panjkov24} & X-ray Model & -- & $< 5 \times 10^{-4}$ & $< 4 \times 10^{15}$ & 50\\
\cite{Nayana25} & X-ray/Radio Model & -- & $10^{-4}$ & $(0.04 - 2) \times 10^{16}$ & 25\\
\cite{Iwata25} & Radio Model & -- & $10^{-6} - 10^{-4}$ & $(0.2 - 1) \times 10^{16}$ & 115\\
\cite{Berger23} & Radio Model & -- & $>10^{-2}$ & $<10^{15}$ & 115\\
\bottomrule
\end{tabularx}
\end{adjustwidth}
\end{table*}


\section{Conclusions}

In this article, we have reviewed the multi-wavelength observations and modeling of type II SN~2023ixf, located at $\sim 6.85$~Mpc in Messier 101. Pre-explosion imaging confirmed that the progenitor star was a dust enshrouded red supergiant, but the exact ZAMS mass remains uncertain. Early-time spectroscopy revealed narrow, transient, high-ionization emission lines that result from SN ejecta interaction with dense, confined CSM that was created by the red supergiant in the final $\sim 3-6$~years before explosion. The most local CSM at $<5 \times 10^{14}$~cm is high density and could be described by a mass loss rate as large as $\sim 10^{-2}$~\mdot. At larger radii, all multi-wavelength (X-ray through radio) observations appear to confirm a similar wind-like density profile with a mass loss rate of $\sim 10^{-4}$~\mdot. Because of the lack of detected outbursts in pre-explosion imaging, mechanisms capable of producing such high density CSM in the final years before explosion could include convection-driven chromosphere \cite{Fuller24, Soker23} and/or binary interaction. Today, SN~2023ixf remains sufficiently bright for detection by ground based instruments and has plateaued in luminosity as shock power dominates over radioactive decay. The unprecedented multi-wavelength dataset of SN~2023ixf has solidified this event as the prototype for CSM-interacting SNe~II, which will continue to be studied for many years to come.

\acknowledgments{We thank Griffin Hosseinzadeh, Charlie Kilpatrick, and A. J. Nayana for contributing figures for this review. W.J.-G. is supported by NASA through Hubble Fellowship grant HSTHF2-51558.001-A awarded by the Space Telescope Science Institute, which is operated for NASA by the Association of Universities for Research in Astronomy, Inc., under contract NAS5-26555. }

\conflictsofinterest{The authors declare no conflicts of interest. } 




\begin{adjustwidth}{-\extralength}{0cm}

\reftitle{References}


\bibliography{references}

\begin{thebibliography}{999}

\bibitem[{Itagaki}(2023)]{Itagaki23}
{Itagaki}, K.
\newblock {Transient Discovery Report for 2023-05-19}.
\newblock {\em Transient Name Server Discovery Report} {\bf 2023}, {\em
  2023-1158},~1.

\bibitem[{Riess} et~al.(2022){Riess}, {Yuan}, {Macri}, {Scolnic}, {Brout},
  {Casertano}, {Jones}, {Murakami}, {Anand}, {Breuval}, {Brink}, {Filippenko},
  {Hoffmann}, {Jha}, {D'arcy Kenworthy}, {Mackenty}, {Stahl}, and
  {Zheng}]{riess22}
{Riess}, A.G.; {Yuan}, W.; {Macri}, L.M.; {Scolnic}, D.; {Brout}, D.;
  {Casertano}, S.; {Jones}, D.O.; {Murakami}, Y.; {Anand}, G.S.; {Breuval}, L.;
   et~al.
\newblock {A Comprehensive Measurement of the Local Value of the Hubble
  Constant with 1 km s$^{-1}$ Mpc$^{-1}$ Uncertainty from the Hubble Space
  Telescope and the SH0ES Team}.
\newblock {\em ApJL} {\bf 2022}, {\em 934},~L7,
  \href{http://arxiv.org/abs/2112.04510}{{\normalfont
  [arXiv:astro-ph.CO/2112.04510]}}.
\newblock {\url{https://doi.org/10.3847/2041-8213/ac5c5b}}.

\bibitem[{Perley} et~al.(2023){Perley}, {Gal-Yam}, {Irani}, and
  {Zimmerman}]{perley23}
{Perley}, D.A.; {Gal-Yam}, A.; {Irani}, I.; {Zimmerman}, E.
\newblock {LT Classification of SN 2023ixf as a Type II Supernova in M101}.
\newblock {\em Transient Name Server AstroNote} {\bf 2023}, {\em 119},~1.

\bibitem[{Sgro} et~al.(2023){Sgro}, {Esposito}, {Blaclard}, {Gomez}, {Marchis},
  {Filippenko}, {Peluso}, {Lawrence}, {Verveen}, {Wagner}, {Nardi}, {Wiart},
  {Mirwald}, {Christensen}, {Eramia}, {Parker}, {Guillet}, {Kim}, {Logan},
  {Kyba}, {Toulmin}, {Vantaggiato}, {Adhis}, {Gary}, {Goodey}, {Dickinson},
  {Koster}, {Martin}, {Bonilla}, {Chung}, {Miny}, {Mortecrette}, {Saibi},
  {Gagnon}, {Simard}, {Vacon}, {Simard}, {Dreise}, {Funakoshi}, {Vacon},
  {Yaniz}, {Le Tarnec}, {Laugier}, {Siders}, {Sweitzer}, {Dvoracek}, {Archer},
  {Deitz}, {Bradley}, {Fukui}, {Sibbernsen}, {Borrot}, {Cross}, {Heider},
  {Yamaguchi}, {Hirsch}, {Leroux}, {Billiani}, {Lorber}, {Smallen}, {Shimizu},
  {Nishimura}, {Ryno}, {Cunningham}, {Gagnon}, {Primm}, {Rushton},
  {Sibbernsen}, {Mitchell}, {Yoblonsky}, {Leroux}, {Clerget}, {Stojanovi{\'c}},
  {Unique}, {Huth}, {Ang}, {Santoni}, {Foster}, {Poggiali}, {Xu}, {Kukita},
  {{\v{S}}{\'c}epanovi{\'c}}, {Saibi}, {Will}, {Latour}, {Haythornthwaite},
  {Cadieux}, {M{\"u}ller}, {Chung}, {Watanabe}, and {Arnaud}]{Sgro23}
{Sgro}, L.A.; {Esposito}, T.M.; {Blaclard}, G.; {Gomez}, S.; {Marchis}, F.;
  {Filippenko}, A.V.; {Peluso}, D.O.; {Lawrence}, S.S.; {Verveen}, A.;
  {Wagner}, A.;  et~al.
\newblock {Photometry of Type II Supernova SN 2023ixf with a Worldwide Citizen
  Science Network}.
\newblock {\em Research Notes of the American Astronomical Society} {\bf 2023},
  {\em 7},~141,  \href{http://arxiv.org/abs/2307.14347}{{\normalfont
  [arXiv:astro-ph.HE/2307.14347]}}.
\newblock {\url{https://doi.org/10.3847/2515-5172/ace41f}}.

\bibitem[{Mao} et~al.(2023){Mao}, {Zhang}, {Cai}, {Chen}, {Chen}, {Gao}, {Li},
  {Lyu}, {Qin}, {Sun}, {Xu}, {Zhang}, {Zhang}, {Zhao}, {Zheng}, {Zhou}, and
  {Ye}]{Mao23}
{Mao}, Y.; {Zhang}, M.; {Cai}, G.; {Chen}, J.; {Chen}, J.; {Gao}, X.; {Li}, K.;
  {Lyu}, X.; {Qin}, Y.; {Sun}, G.;  et~al.
\newblock {Onset of SN 2023ixf observed over East Asian longitudes}.
\newblock {\em Transient Name Server AstroNote} {\bf 2023}, {\em 130},~1.

\bibitem[{Li} et~al.(2024){Li}, {Hu}, {Li}, {Yang}, {Wang}, {Yan}, {Hu},
  {Zhang}, {Mao}, {Riise}, {Gao}, {Sun}, {Liu}, {Xiong}, {Wang}, {Mo},
  {Iskandar}, {Xi}, {Xiang}, {Wang}, {Sun}, {Zhang}, {Chen}, {Lin}, {Guo},
  {Liu}, {Cai}, {Zhou}, {Zhao}, {Chen}, {Zheng}, {Li}, {Zhang}, {Xu}, {Lyu},
  {Castro-Tirado}, {Chufarin}, {Potapov}, {Ionov}, {Korotkiy}, {Nazarov},
  {Sokolovsky}, {Hamann}, and {Herman}]{Li24}
{Li}, G.; {Hu}, M.; {Li}, W.; {Yang}, Y.; {Wang}, X.; {Yan}, S.; {Hu}, L.;
  {Zhang}, J.; {Mao}, Y.; {Riise}, H.;  et~al.
\newblock {A shock flash breaking out of a dusty red supergiant}.
\newblock {\em Nature} {\bf 2024}, {\em 627},~754--758,
  \href{http://arxiv.org/abs/2311.14409}{{\normalfont
  [arXiv:astro-ph.HE/2311.14409]}}.
\newblock {\url{https://doi.org/10.1038/s41586-023-06843-6}}.

\bibitem[{Hosseinzadeh} et~al.(2023){Hosseinzadeh}, {Farah}, {Shrestha},
  {Sand}, {Dong}, {Brown}, {Bostroem}, {Valenti}, {Jha}, {Andrews}, {Arcavi},
  {Haislip}, {Hiramatsu}, {Hoang}, {Howell}, {Janzen}, {Jencson}, {Kouprianov},
  {Lundquist}, {McCully}, {Meza Retamal}, {Modjaz}, {Newsome}, {Padilla
  Gonzalez}, {Pearson}, {Pellegrino}, {Ravi}, {Reichart}, {Smith}, {Terreran},
  and {Vink{\'o}}]{Hosseinzadeh23}
{Hosseinzadeh}, G.; {Farah}, J.; {Shrestha}, M.; {Sand}, D.J.; {Dong}, Y.;
  {Brown}, P.J.; {Bostroem}, K.A.; {Valenti}, S.; {Jha}, S.W.; {Andrews}, J.E.;
   et~al.
\newblock {Shock Cooling and Possible Precursor Emission in the Early Light
  Curve of the Type II SN 2023ixf}.
\newblock {\em ApJL} {\bf 2023}, {\em 953},~L16,
  \href{http://arxiv.org/abs/2306.06097}{{\normalfont
  [arXiv:astro-ph.HE/2306.06097]}}.
\newblock {\url{https://doi.org/10.3847/2041-8213/ace4c4}}.

\bibitem[{Jacobson-Gal{\'a}n} et~al.(2023){Jacobson-Gal{\'a}n}, {Dessart},
  {Margutti}, {Chornock}, {Foley}, {Kilpatrick}, {Jones}, {Taggart}, {Angus},
  {Bhattacharjee}, {Braff}, {Brethauer}, {Burgasser}, {Cao}, {Carlile},
  {Chambers}, {Coulter}, {Dominguez-Ruiz}, {Dickinson}, {de Boer}, {Gagliano},
  {Gall}, {Gao}, {Gates}, {Gomez}, {Guolo}, {Halford}, {Hjorth}, {Huber},
  {Johnson}, {Karpoor}, {Laskar}, {LeBaron}, {Li}, {Lin}, {Loch}, {Lynam},
  {Magnier}, {Maloney}, {Matthews}, {McDonald}, {Miao}, {Milisavljevic}, {Pan},
  {Pradyumna}, {Ransome}, {Rees}, {Rest}, {Rojas-Bravo}, {Sandford},
  {Ascencio}, {Sanjaripour}, {Savino}, {Sears}, {Sharei}, {Smartt}, {Softich},
  {Theissen}, {Tinyanont}, {Tohfa}, {Villar}, {Wang}, {Wainscoat},
  {Westerling}, {Wiston}, {Wozniak}, {Yadavalli}, and {Zenati}]{wjg23}
{Jacobson-Gal{\'a}n}, W.V.; {Dessart}, L.; {Margutti}, R.; {Chornock}, R.;
  {Foley}, R.J.; {Kilpatrick}, C.D.; {Jones}, D.O.; {Taggart}, K.; {Angus},
  C.R.; {Bhattacharjee}, S.;  et~al.
\newblock {SN 2023ixf in Messier 101: Photo-ionization of Dense, Close-in
  Circumstellar Material in a Nearby Type II Supernova}.
\newblock {\em ApJL} {\bf 2023}, {\em 954},~L42,
  \href{http://arxiv.org/abs/2306.04721}{{\normalfont
  [arXiv:astro-ph.HE/2306.04721]}}.
\newblock {\url{https://doi.org/10.3847/2041-8213/acf2ec}}.

\bibitem[{Bostroem} et~al.(2023){Bostroem}, {Pearson}, {Shrestha}, {Sand},
  {Valenti}, {Jha}, {Andrews}, {Smith}, {Terreran}, {Green}, {Dong},
  {Lundquist}, {Haislip}, {Hoang}, {Hosseinzadeh}, {Janzen}, {Jencson},
  {Kouprianov}, {Paraskeva}, {Meza Retamal}, {Reichart}, {Arcavi}, {Bonanos},
  {Coughlin}, {Dobson}, {Farah}, {Galbany}, {Guti{\'e}rrez}, {Hawley}, {Hebb},
  {Hiramatsu}, {Howell}, {Iijima}, {Ilyin}, {Jhass}, {McCully}, {Moran},
  {Morris}, {Mura}, {M{\"u}ller-Bravo}, {Munday}, {Newsome}, {Pabst}, {Ochner},
  {Gonzalez}, {Pastorello}, {Pellegrino}, {Piscarreta}, {Ravi}, {Reguitti},
  {Salo}, {Vink{\'o}}, {de Vos}, {Wheeler}, {Williams}, and
  {Wyatt}]{Bostroem23}
{Bostroem}, K.A.; {Pearson}, J.; {Shrestha}, M.; {Sand}, D.J.; {Valenti}, S.;
  {Jha}, S.W.; {Andrews}, J.E.; {Smith}, N.; {Terreran}, G.; {Green}, E.;
  et~al.
\newblock {Early Spectroscopy and Dense Circumstellar Medium Interaction in SN
  2023ixf}.
\newblock {\em ApJL} {\bf 2023}, {\em 956},~L5,
  \href{http://arxiv.org/abs/2306.10119}{{\normalfont
  [arXiv:astro-ph.HE/2306.10119]}}.
\newblock {\url{https://doi.org/10.3847/2041-8213/acf9a4}}.

\bibitem[{Teja} et~al.(2023){Teja}, {Singh}, {Basu}, {Anupama}, {Sahu},
  {Dutta}, {Swain}, {Nakaoka}, {Pathak}, {Bhalerao}, {Barway}, {Kumar},
  {A.~J.}, {Imazawa}, {Kumar}, and {Kawabata}]{Teja23}
{Teja}, R.S.; {Singh}, A.; {Basu}, J.; {Anupama}, G.C.; {Sahu}, D.K.; {Dutta},
  A.; {Swain}, V.; {Nakaoka}, T.; {Pathak}, U.; {Bhalerao}, V.;  et~al.
\newblock {Far-ultraviolet to Near-infrared Observations of SN 2023ixf: A
  High-energy Explosion Engulfed in Complex Circumstellar Material}.
\newblock {\em ApJL} {\bf 2023}, {\em 954},~L12,
  \href{http://arxiv.org/abs/2306.10284}{{\normalfont
  [arXiv:astro-ph.HE/2306.10284]}}.
\newblock {\url{https://doi.org/10.3847/2041-8213/acef20}}.

\bibitem[{Zhang} et~al.(2023){Zhang}, {Lin}, {Wang}, {Zhao}, {Li}, {Liu},
  {Yan}, {Xiang}, {Wang}, and {Bai}]{Zhang23}
{Zhang}, J.; {Lin}, H.; {Wang}, X.; {Zhao}, Z.; {Li}, L.; {Liu}, J.; {Yan}, S.;
  {Xiang}, D.; {Wang}, H.; {Bai}, J.
\newblock {Circumstellar material ejected violently by a massive star
  immediately before its death}.
\newblock {\em Science Bulletin} {\bf 2023}, {\em 68},~2548--2554,
  \href{http://arxiv.org/abs/2309.01998}{{\normalfont
  [arXiv:astro-ph.HE/2309.01998]}}.
\newblock {\url{https://doi.org/10.1016/j.scib.2023.09.015}}.

\bibitem[{Gal-Yam} et~al.(2014){Gal-Yam}, {Arcavi}, {Ofek}, {Ben-Ami}, {Cenko},
  {Kasliwal}, {Cao}, {Yaron}, {Tal}, {Silverman}, {Horesh}, {De Cia}, {Taddia},
  {Sollerman}, {Perley}, {Vreeswijk}, {Kulkarni}, {Nugent}, {Filippenko}, and
  {Wheeler}]{galyam14}
{Gal-Yam}, A.; {Arcavi}, I.; {Ofek}, E.O.; {Ben-Ami}, S.; {Cenko}, S.B.;
  {Kasliwal}, M.M.; {Cao}, Y.; {Yaron}, O.; {Tal}, D.; {Silverman}, J.M.;
  et~al.
\newblock {A Wolf-Rayet-like progenitor of SN 2013cu from spectral observations
  of a stellar wind}.
\newblock {\em Nature} {\bf 2014}, {\em 509},~471--474,
  \href{http://arxiv.org/abs/1406.7640}{{\normalfont
  [arXiv:astro-ph.HE/1406.7640]}}.
\newblock {\url{https://doi.org/10.1038/nature13304}}.

\bibitem[{Yaron} et~al.(2017){Yaron}, {Perley}, {Gal-Yam}, {Groh}, {Horesh},
  {Ofek}, {Kulkarni}, {Sollerman}, {Fransson}, {Rubin}, {Szabo}, {Sapir},
  {Taddia}, {Cenko}, {Valenti}, {Arcavi}, {Howell}, {Kasliwal}, {Vreeswijk},
  {Khazov}, {Fox}, {Cao}, {Gnat}, {Kelly}, {Nugent}, {Filippenko}, {Laher},
  {Wozniak}, {Lee}, {Rebbapragada}, {Maguire}, {Sullivan}, and
  {Soumagnac}]{yaron17}
{Yaron}, O.; {Perley}, D.A.; {Gal-Yam}, A.; {Groh}, J.H.; {Horesh}, A.; {Ofek},
  E.O.; {Kulkarni}, S.R.; {Sollerman}, J.; {Fransson}, C.; {Rubin}, A.;  et~al.
\newblock {Confined dense circumstellar material surrounding a regular type II
  supernova}.
\newblock {\em Nature Physics} {\bf 2017}, {\em 13},~510--517,
  \href{http://arxiv.org/abs/1701.02596}{{\normalfont
  [arXiv:astro-ph.HE/1701.02596]}}.
\newblock {\url{https://doi.org/10.1038/nphys4025}}.

\bibitem[{Dessart} et~al.(2017){Dessart}, {John Hillier}, and
  {Audit}]{dessart17}
{Dessart}, L.; {John Hillier}, D.; {Audit}, E.
\newblock {Explosion of red-supergiant stars: Influence of the atmospheric
  structure on shock breakout and early-time supernova radiation}.
\newblock {\em A\&A} {\bf 2017}, {\em 605},~A83,
  \href{http://arxiv.org/abs/1704.01697}{{\normalfont
  [arXiv:astro-ph.SR/1704.01697]}}.
\newblock {\url{https://doi.org/10.1051/0004-6361/201730942}}.

\bibitem[{Khazov} et~al.(2016){Khazov}, {Yaron}, {Gal-Yam}, {Manulis}, {Rubin},
  {Kulkarni}, {Arcavi}, {Kasliwal}, {Ofek}, {Cao}, {Perley}, {Sollerman},
  {Horesh}, {Sullivan}, {Filippenko}, {Nugent}, {Howell}, {Cenko}, {Silverman},
  {Ebeling}, {Taddia}, {Johansson}, {Laher}, {Surace}, {Rebbapragada},
  {Wozniak}, and {Matheson}]{Khazov16}
{Khazov}, D.; {Yaron}, O.; {Gal-Yam}, A.; {Manulis}, I.; {Rubin}, A.;
  {Kulkarni}, S.R.; {Arcavi}, I.; {Kasliwal}, M.M.; {Ofek}, E.O.; {Cao}, Y.;
  et~al.
\newblock {Flash Spectroscopy: Emission Lines from the Ionized Circumstellar
  Material around <10-day-old Type II Supernovae}.
\newblock {\em ApJ} {\bf 2016}, {\em 818},~3,
  \href{http://arxiv.org/abs/1512.00846}{{\normalfont
  [arXiv:astro-ph.HE/1512.00846]}}.
\newblock {\url{https://doi.org/10.3847/0004-637X/818/1/3}}.

\bibitem[{Bruch} et~al.(2021){Bruch}, {Gal-Yam}, {Schulze}, {Yaron}, {Yang},
  {Soumagnac}, {Rigault}, {Strotjohann}, {Ofek}, {Sollerman}, {Masci},
  {Barbarino}, {Ho}, {Fremling}, {Perley}, {Nordin}, {Cenko}, {Adams},
  {Adreoni}, {Bellm}, {Blagorodnova}, {Bulla}, {Burdge}, {De}, {Dhawan},
  {Drake}, {Duev}, {Dugas}, {Graham}, {Graham}, {Irani}, {Jencson},
  {Karamehmetoglu}, {Kasliwal}, {Kim}, {Kulkarni}, {Kupfer}, {Liang},
  {Mahabal}, {Miller}, {Prince}, {Riddle}, {Sharma}, {Smith}, {Taddia},
  {Taggart}, {Walters}, and {Yan}]{bruch21}
{Bruch}, R.J.; {Gal-Yam}, A.; {Schulze}, S.; {Yaron}, O.; {Yang}, Y.;
  {Soumagnac}, M.; {Rigault}, M.; {Strotjohann}, N.L.; {Ofek}, E.; {Sollerman},
  J.;  et~al.
\newblock {A Large Fraction of Hydrogen-rich Supernova Progenitors Experience
  Elevated Mass Loss Shortly Prior to Explosion}.
\newblock {\em ApJ} {\bf 2021}, {\em 912},~46,
  \href{http://arxiv.org/abs/2008.09986}{{\normalfont
  [arXiv:astro-ph.HE/2008.09986]}}.
\newblock {\url{https://doi.org/10.3847/1538-4357/abef05}}.

\bibitem[{Bruch} et~al.(2023){Bruch}, {Gal-Yam}, {Yaron}, {Chen},
  {Strotjohann}, {Irani}, {Zimmerman}, {Schulze}, {Yang}, {Kim}, {Bulla},
  {Sollerman}, {Rigault}, {Ofek}, {Soumagnac}, {Masci}, {Fremling}, {Perley},
  {Nordin}, {Cenko}, {Ho}, {Adams}, {Adreoni}, {Bellm}, {Blagorodnova},
  {Burdge}, {De}, {Dekany}, {Dhawan}, {Drake}, {Duev}, {Graham}, {Graham},
  {Jencson}, {Karamehmetoglu}, {Kasliwal}, {Kulkarni}, {Miller}, {Neill},
  {Prince}, {Riddle}, {Rusholme}, {Sharma}, {Smith}, {Sravan}, {Taggart},
  {Walters}, and {Yan}]{bruch23}
{Bruch}, R.J.; {Gal-Yam}, A.; {Yaron}, O.; {Chen}, P.; {Strotjohann}, N.L.;
  {Irani}, I.; {Zimmerman}, E.; {Schulze}, S.; {Yang}, Y.; {Kim}, Y.L.;  et~al.
\newblock {The Prevalence and Influence of Circumstellar Material around
  Hydrogen-rich Supernova Progenitors}.
\newblock {\em ApJ} {\bf 2023}, {\em 952},~119,
  \href{http://arxiv.org/abs/2212.03313}{{\normalfont
  [arXiv:astro-ph.HE/2212.03313]}}.
\newblock {\url{https://doi.org/10.3847/1538-4357/acd8be}}.

\bibitem[{Chugai}(2001)]{Chugai01}
{Chugai}, N.N.
\newblock {Broad emission lines from the opaque electron-scattering environment
  of SN 1998S}.
\newblock {\em MNRAS} {\bf 2001}, {\em 326},~1448--1454,
  \href{http://arxiv.org/abs/astro-ph/0106234}{{\normalfont
  [arXiv:astro-ph/astro-ph/0106234]}}.
\newblock {\url{https://doi.org/10.1111/j.1365-2966.2001.04717.x}}.

\bibitem[{Huang} and {Chevalier}(2018)]{Huang18}
{Huang}, C.; {Chevalier}, R.A.
\newblock {Electron scattering wings on lines in interacting supernovae}.
\newblock {\em MNRAS} {\bf 2018}, {\em 475},~1261--1273,
  \href{http://arxiv.org/abs/1712.01237}{{\normalfont
  [arXiv:astro-ph.HE/1712.01237]}}.
\newblock {\url{https://doi.org/10.1093/mnras/stx3163}}.

\bibitem[{Dessart}(2025)]{Dessart25}
{Dessart}, L.
\newblock {Probing red supergiant atmospheres and winds with early-time,
  high-cadence, high-resolution type II supernova spectra}.
\newblock {\em A\&A} {\bf 2025}, {\em 694},~A132,
  \href{http://arxiv.org/abs/2410.20486}{{\normalfont
  [arXiv:astro-ph.HE/2410.20486]}}.
\newblock {\url{https://doi.org/10.1051/0004-6361/202452769}}.

\bibitem[{Dickinson} et~al.(2025){Dickinson}, {Milisavljevic}, {Garretson},
  {Dessart}, {Margutti}, {Chornock}, {Subrayan}, {Hillier}, {Golub}, {Li},
  {Logsdon}, {Rajagopal}, {Ridgway}, {Smith}, and {Cynamon}]{Dickinson25}
{Dickinson}, D.; {Milisavljevic}, D.; {Garretson}, B.; {Dessart}, L.;
  {Margutti}, R.; {Chornock}, R.; {Subrayan}, B.; {Hillier}, D.J.; {Golub}, E.;
  {Li}, D.;  et~al.
\newblock {The Immediate, Exemplary, and Fleeting Echelle Spectroscopy of SN
  2023ixf: Monitoring Acceleration of Slow Progenitor Circumstellar Material
  Driven by Shock Interaction}.
\newblock {\em ApJ} {\bf 2025}, {\em 984},~71,
  \href{http://arxiv.org/abs/2412.14406}{{\normalfont
  [arXiv:astro-ph.HE/2412.14406]}}.
\newblock {\url{https://doi.org/10.3847/1538-4357/adc108}}.

\bibitem[{Shivvers} et~al.(2015){Shivvers}, {Groh}, {Mauerhan}, {Fox},
  {Leonard}, and {Filippenko}]{shivvers15}
{Shivvers}, I.; {Groh}, J.H.; {Mauerhan}, J.C.; {Fox}, O.D.; {Leonard}, D.C.;
  {Filippenko}, A.V.
\newblock {Early Emission from the Type IIn Supernova 1998S at High
  Resolution}.
\newblock {\em ApJ} {\bf 2015}, {\em 806},~213,
  \href{http://arxiv.org/abs/1408.1404}{{\normalfont
  [arXiv:astro-ph.HE/1408.1404]}}.
\newblock {\url{https://doi.org/10.1088/0004-637X/806/2/213}}.

\bibitem[{Smith}(2017)]{Smith17}
{Smith}, N.
\newblock {Interacting Supernovae: Types IIn and Ibn}. In {\em Handbook of
  Supernovae}; {Alsabti}, A.W.; {Murdin}, P., Eds.;  2017; p. 403.
\newblock {\url{https://doi.org/10.1007/978-3-319-21846-5_38}}.

\bibitem[{Smith} et~al.(2023){Smith}, {Pearson}, {Sand}, {Ilyin}, {Bostroem},
  {Hosseinzadeh}, and {Shrestha}]{Smith23}
{Smith}, N.; {Pearson}, J.; {Sand}, D.J.; {Ilyin}, I.; {Bostroem}, K.A.;
  {Hosseinzadeh}, G.; {Shrestha}, M.
\newblock {High-resolution Spectroscopy of SN 2023ixf's First Week: Engulfing
  the Asymmetric Circumstellar Material}.
\newblock {\em ApJ} {\bf 2023}, {\em 956},~46,
  \href{http://arxiv.org/abs/2306.07964}{{\normalfont
  [arXiv:astro-ph.HE/2306.07964]}}.
\newblock {\url{https://doi.org/10.3847/1538-4357/acf366}}.

\bibitem[{Vasylyev} et~al.(2023){Vasylyev}, {Yang}, {Filippenko}, {Patra},
  {Brink}, {Wang}, {Chornock}, {Margutti}, {Gates}, {Burgasser}, {Karpoor},
  {LeBaron}, {Softich}, {Theissen}, {Wiston}, and {Zheng}]{Vasylyev23}
{Vasylyev}, S.S.; {Yang}, Y.; {Filippenko}, A.V.; {Patra}, K.C.; {Brink}, T.G.;
  {Wang}, L.; {Chornock}, R.; {Margutti}, R.; {Gates}, E.L.; {Burgasser}, A.J.;
   et~al.
\newblock {Early Time Spectropolarimetry of the Aspherical Type II Supernova SN
  2023ixf}.
\newblock {\em ApJL} {\bf 2023}, {\em 955},~L37,
  \href{http://arxiv.org/abs/2307.01268}{{\normalfont
  [arXiv:astro-ph.HE/2307.01268]}}.
\newblock {\url{https://doi.org/10.3847/2041-8213/acf1a3}}.

\bibitem[{Vasylyev} et~al.(2025){Vasylyev}, {Dessart}, {Yang}, {Filippenko},
  {Patra}, {Brink}, {Wang}, {Chornock}, {Margutti}, {Gates}, {Burgasser},
  {Sears}, {Karpoor}, {LeBaron}, {Softich}, {Theissen}, {Wiston}, and
  {Zheng}]{Vasylyev25}
{Vasylyev}, S.S.; {Dessart}, L.; {Yang}, Y.; {Filippenko}, A.V.; {Patra}, K.C.;
  {Brink}, T.G.; {Wang}, L.; {Chornock}, R.; {Margutti}, R.; {Gates}, E.L.;
  et~al.
\newblock {Spectropolarimetric Evolution of SN 2023ixf: an Asymmetric Explosion
  in a Confined Aspherical Circumstellar Medium}.
\newblock {\em arXiv e-prints} {\bf 2025}, p. arXiv:2505.03975,
  \href{http://arxiv.org/abs/2505.03975}{{\normalfont
  [arXiv:astro-ph.HE/2505.03975]}}.
\newblock {\url{https://doi.org/10.48550/arXiv.2505.03975}}.

\bibitem[{Jacobson-Gal{\'a}n} et~al.(2024){Jacobson-Gal{\'a}n}, {Dessart},
  {Davis}, {Kilpatrick}, {Margutti}, {Foley}, {Chornock}, {Terreran},
  {Hiramatsu}, {Newsome}, {Padilla Gonzalez}, {Pellegrino}, {Howell},
  {Filippenko}, {Anderson}, {Angus}, {Auchettl}, {Bostroem}, {Brink},
  {Cartier}, {Coulter}, {de Boer}, {Drout}, {Earl}, {Ertini}, {Farah},
  {Farias}, {Gall}, {Gao}, {Gerlach}, {Guo}, {Haynie}, {Hosseinzadeh}, {Ibik},
  {Jha}, {Jones}, {Langeroodi}, {LeBaron}, {Magnier}, {Piro}, {Raimundo},
  {Rest}, {Rest}, {Rich}, {Rojas-Bravo}, {Sears}, {Taggart}, {Villar},
  {Wainscoat}, {Wang}, {Wasserman}, {Yan}, {Yang}, {Zhang}, and
  {Zheng}]{wjg24a}
{Jacobson-Gal{\'a}n}, W.V.; {Dessart}, L.; {Davis}, K.W.; {Kilpatrick}, C.D.;
  {Margutti}, R.; {Foley}, R.J.; {Chornock}, R.; {Terreran}, G.; {Hiramatsu},
  D.; {Newsome}, M.;  et~al.
\newblock {Final Moments. II. Observational Properties and Physical Modeling of
  Circumstellar-material-interacting Type II Supernovae}.
\newblock {\em ApJ} {\bf 2024}, {\em 970},~189,
  \href{http://arxiv.org/abs/2403.02382}{{\normalfont
  [arXiv:astro-ph.HE/2403.02382]}}.
\newblock {\url{https://doi.org/10.3847/1538-4357/ad4a2a}}.

\bibitem[{Yamanaka} et~al.(2023){Yamanaka}, {Fujii}, and
  {Nagayama}]{Yamanaka23}
{Yamanaka}, M.; {Fujii}, M.; {Nagayama}, T.
\newblock {Bright Type II supernova 2023ixf in M 101: A quick analysis of the
  early-stage spectra and near-infrared light curves}.
\newblock {\em PASJ} {\bf 2023}, {\em 75},~L27--L31,
  \href{http://arxiv.org/abs/2306.00263}{{\normalfont
  [arXiv:astro-ph.SR/2306.00263]}}.
\newblock {\url{https://doi.org/10.1093/pasj/psad051}}.

\bibitem[{Zimmerman} et~al.(2024){Zimmerman}, {Irani}, {Chen}, {Gal-Yam},
  {Schulze}, {Perley}, {Sollerman}, {Filippenko}, {Shenar}, {Yaron}, {Shahaf},
  {Bruch}, {Ofek}, {De Cia}, {Brink}, {Yang}, {Vasylyev}, {Ben Ami}, {Aubert},
  {Badash}, {Bloom}, {Brown}, {De}, {Dimitriadis}, {Fransson}, {Fremling},
  {Hinds}, {Horesh}, {Johansson}, {Kasliwal}, {Kulkarni}, {Kushnir}, {Martin},
  {Matuzewski}, {McGurk}, {Miller}, {Morag}, {Neil}, {Nugent}, {Post},
  {Prusinski}, {Qin}, {Raichoor}, {Riddle}, {Rowe}, {Rusholme}, {Sfaradi},
  {Sjoberg}, {Soumagnac}, {Stein}, {Strotjohann}, {Terwel}, {Wasserman},
  {Wise}, {Wold}, {Yan}, and {Zhang}]{Zimmerman24}
{Zimmerman}, E.A.; {Irani}, I.; {Chen}, P.; {Gal-Yam}, A.; {Schulze}, S.;
  {Perley}, D.A.; {Sollerman}, J.; {Filippenko}, A.V.; {Shenar}, T.; {Yaron},
  O.;  et~al.
\newblock {The complex circumstellar environment of supernova 2023ixf}.
\newblock {\em Nature} {\bf 2024}, {\em 627},~759--762,
  \href{http://arxiv.org/abs/2310.10727}{{\normalfont
  [arXiv:astro-ph.HE/2310.10727]}}.
\newblock {\url{https://doi.org/10.1038/s41586-024-07116-6}}.

\bibitem[{Dessart} and {Jacobson-Gal{\'a}n}(2023)]{Dessart23}
{Dessart}, L.; {Jacobson-Gal{\'a}n}, W.V.
\newblock {Using spectral modeling to break light-curve degeneracies of type II
  supernovae interacting with circumstellar material}.
\newblock {\em A\&A} {\bf 2023}, {\em 677},~A105,
  \href{http://arxiv.org/abs/2307.08584}{{\normalfont
  [arXiv:astro-ph.SR/2307.08584]}}.
\newblock {\url{https://doi.org/10.1051/0004-6361/202346754}}.

\bibitem[{Jacobson-Gal{\'a}n} et~al.(2024){Jacobson-Gal{\'a}n}, {Davis},
  {Kilpatrick}, {Dessart}, {Margutti}, {Chornock}, {Foley}, {Arunachalam},
  {Auchettl}, {Bom}, {Cartier}, {Coulter}, {Dimitriadis}, {Dickinson}, {Drout},
  {Gagliano}, {Gall}, {Garretson}, {Izzo}, {Jones}, {LeBaron}, {Miao},
  {Milisavljevic}, {Pan}, {Rest}, {Rojas-Bravo}, {Santos}, {Sears}, {Subrayan},
  {Taggart}, and {Tinyanont}]{wjg24b}
{Jacobson-Gal{\'a}n}, W.V.; {Davis}, K.W.; {Kilpatrick}, C.D.; {Dessart}, L.;
  {Margutti}, R.; {Chornock}, R.; {Foley}, R.J.; {Arunachalam}, P.; {Auchettl},
  K.; {Bom}, C.R.;  et~al.
\newblock {SN 2024ggi in NGC 3621: Rising Ionization in a Nearby,
  Circumstellar-material-interacting Type II Supernova}.
\newblock {\em ApJ} {\bf 2024}, {\em 972},~177,
  \href{http://arxiv.org/abs/2404.19006}{{\normalfont
  [arXiv:astro-ph.HE/2404.19006]}}.
\newblock {\url{https://doi.org/10.3847/1538-4357/ad5c64}}.

\bibitem[{Zhang} et~al.(2024){Zhang}, {Li}, {Cheng}, {Wu}, {Jia}, {Chen},
  {Cui}, {Feng}, {Guan}, {Han}, {Li}, {Liu}, {Lu}, {Song}, {Wang}, {Xu},
  {Zhang}, {Zhao}, {Zhao}, {Jin}, {Ling}, {Liu}, {Liu}, {Liu}, {Li}, {Sun},
  {Yuan}, {Zhang}, {Zhang}, {Li}, {Wang}, {Zhou}, {Nandra}, {Rau}, {Friedrich},
  {Meidinger}, {Burwitz}, {Kuulkers}, {Santovincenzo}, {O'Brien}, {Cordier},
  {Wang}, and {Li}]{zhang24}
{Zhang}, J.; {Li}, C.K.; {Cheng}, H.Q.; {Wu}, Q.Y.; {Jia}, S.M.; {Chen}, Y.;
  {Cui}, W.W.; {Feng}, H.; {Guan}, J.; {Han}, D.W.;  et~al.
\newblock {SN 2024ggi: detection of X-ray emission by EP-FXT}.
\newblock {\em The Astronomer's Telegram} {\bf 2024}, {\em 16588},~1.

\bibitem[{Kulkarni} et~al.(2021){Kulkarni}, {Harrison}, {Grefenstette},
  {Earnshaw}, {Andreoni}, {Berg}, {Bloom}, {Cenko}, {Chornock}, {Christiansen},
  {Coughlin}, {Wuollet Criswell}, {Darvish}, {Das}, {De}, {Dessart}, {Dixon},
  {Dorsman}, {El-Badry}, {Evans}, {Ford}, {Fremling}, {Gansicke}, {Gezari},
  {Goetberg}, {Green}, {Graham}, {Heida}, {Ho}, {Jaodand}, {Johns-Krull},
  {Kasliwal}, {Lazzarini}, {Lu}, {Margutti}, {Martin}, {Masters}, {McKernan},
  {Naze}, {Nissanke}, {Parazin}, {Perley}, {Phinney}, {Piro}, {Raaijmakers},
  {Rauw}, {Rodriguez}, {Sana}, {Senchyna}, {Singer}, {Spake}, {Stassun},
  {Stern}, {Teplitz}, {Weisz}, and {Yao}]{Kulkarni21}
{Kulkarni}, S.R.; {Harrison}, F.A.; {Grefenstette}, B.W.; {Earnshaw}, H.P.;
  {Andreoni}, I.; {Berg}, D.A.; {Bloom}, J.S.; {Cenko}, S.B.; {Chornock}, R.;
  {Christiansen}, J.L.;  et~al.
\newblock {Science with the Ultraviolet Explorer (UVEX)}.
\newblock {\em arXiv e-prints} {\bf 2021}, p. arXiv:2111.15608,
  \href{http://arxiv.org/abs/2111.15608}{{\normalfont
  [arXiv:astro-ph.GA/2111.15608]}}.
\newblock {\url{https://doi.org/10.48550/arXiv.2111.15608}}.

\bibitem[{Singh} et~al.(2024){Singh}, {Teja}, {Moriya}, {Maeda}, {Kawabata},
  {Tanaka}, {Imazawa}, {Nakaoka}, {Gangopadhyay}, {Yamanaka}, {Swain}, {Sahu},
  {Anupama}, {Kumar}, {Anche}, {Sano}, {Raj}, {Agnihotri}, {Bhalerao}, {Bisht},
  {Bisht}, {Belwal}, {Chakrabarti}, {Fujii}, {Nagayama}, {Matsumoto}, {Hamada},
  {Kawabata}, {Kumar}, {Kumar}, {Malkan}, {Smith}, {Sakagami}, {Taguchi},
  {Tominaga}, and {Watanabe}]{Singh24}
{Singh}, A.; {Teja}, R.S.; {Moriya}, T.J.; {Maeda}, K.; {Kawabata}, K.S.;
  {Tanaka}, M.; {Imazawa}, R.; {Nakaoka}, T.; {Gangopadhyay}, A.; {Yamanaka},
  M.;  et~al.
\newblock {Unravelling the Asphericities in the Explosion and Multifaceted
  Circumstellar Matter of SN 2023ixf}.
\newblock {\em ApJ} {\bf 2024}, {\em 975},~132,
  \href{http://arxiv.org/abs/2405.20989}{{\normalfont
  [arXiv:astro-ph.HE/2405.20989]}}.
\newblock {\url{https://doi.org/10.3847/1538-4357/ad7955}}.

\bibitem[{Chugai} et~al.(2007){Chugai}, {Chevalier}, and {Utrobin}]{Chugai07}
{Chugai}, N.N.; {Chevalier}, R.A.; {Utrobin}, V.P.
\newblock {Optical Signatures of Circumstellar Interaction in Type IIP
  Supernovae}.
\newblock {\em ApJ} {\bf 2007}, {\em 662},~1136--1147,
  \href{http://arxiv.org/abs/astro-ph/0703468}{{\normalfont
  [arXiv:astro-ph/astro-ph/0703468]}}.
\newblock {\url{https://doi.org/10.1086/518160}}.

\bibitem[{Dessart} and {Hillier}(2022)]{Dessart22}
{Dessart}, L.; {Hillier}, D.J.
\newblock {Modeling the signatures of interaction in Type II supernovae: UV
  emission, high-velocity features, broad-boxy profiles}.
\newblock {\em A\&A} {\bf 2022}, {\em 660},~L9,
  \href{http://arxiv.org/abs/2204.00446}{{\normalfont
  [arXiv:astro-ph.SR/2204.00446]}}.
\newblock {\url{https://doi.org/10.1051/0004-6361/202243372}}.

\bibitem[{Zheng} et~al.(2025){Zheng}, {Dessart}, {Filippenko}, {Yang}, {Brink},
  {De Jaeger}, {Vasylyev}, {Van Dyk}, {Patra}, {Jacobson-Galan}, {Stewart},
  {Alvarado}, {Arikatla}, {Beddow}, {Betz}, {Born}, {Bostow}, {Burgasser},
  {Caceres}, {Carrasco}, {Chuang}, {DeGraw}, {Gates}, {Gendreau-Distler},
  {Jacobus}, {Jennings}, {Karpoor}, {Lynam}, {Mina}, {Mora}, {Pichay}, {Ravi},
  {Rees}, {Rich}, {Risin}, {Sandford}, {Savino}, {Softich}, {Theissen},
  {Vidal}, {Wu}, and {Zeng}]{Zheng25}
{Zheng}, W.; {Dessart}, L.; {Filippenko}, A.V.; {Yang}, Y.; {Brink}, T.G.; {De
  Jaeger}, T.; {Vasylyev}, S.S.; {Van Dyk}, S.D.; {Patra}, K.C.;
  {Jacobson-Galan}, W.V.;  et~al.
\newblock {SN 2023ixf in the Pinwheel Galaxy M101: From Shock Breakout to the
  Nebular Phase}.
\newblock {\em arXiv e-prints} {\bf 2025}, p. arXiv:2503.13974,
  \href{http://arxiv.org/abs/2503.13974}{{\normalfont
  [arXiv:astro-ph.GA/2503.13974]}}.
\newblock {\url{https://doi.org/10.48550/arXiv.2503.13974}}.

\bibitem[{Bostroem} et~al.(2024){Bostroem}, {Sand}, {Dessart}, {Smith}, {Jha},
  {Valenti}, {Andrews}, {Dong}, {Filippenko}, {Gomez}, {Hiramatsu}, {Hoang},
  {Hosseinzadeh}, {Howell}, {Jencson}, {Lundquist}, {McCully}, {Mehta},
  {Meza-Retamal}, {Pearson}, {Ravi}, {Shrestha}, and {Wyatt}]{Bostroem24}
{Bostroem}, K.A.; {Sand}, D.J.; {Dessart}, L.; {Smith}, N.; {Jha}, S.W.;
  {Valenti}, S.; {Andrews}, J.E.; {Dong}, Y.; {Filippenko}, A.V.; {Gomez}, S.;
  et~al.
\newblock {Circumstellar Interaction in the Ultraviolet Spectra of SN 2023ixf
  14{\textendash}66 Days After Explosion}.
\newblock {\em ApJL} {\bf 2024}, {\em 973},~L47,
  \href{http://arxiv.org/abs/2408.03993}{{\normalfont
  [arXiv:astro-ph.HE/2408.03993]}}.
\newblock {\url{https://doi.org/10.3847/2041-8213/ad7855}}.

\bibitem[{Shrestha} et~al.(2025){Shrestha}, {DeSoto}, {Sand}, {Williams},
  {Hoffman}, {Smith}, {McCall}, {Maund}, {Steele}, {Wiersema}, {Andrews},
  {Smith}, {Bilinski}, {Milne}, {Anche}, {Bostroem}, {Hosseinzadeh}, {Pearson},
  {Leonard}, {Hsu}, {Dong}, {Hoang}, {Janzen}, {Jencson}, {Jha}, {Lundquist},
  {Mehta}, {Retamal}, {Valenti}, {Farah}, {Howell}, {McCully}, {Newsome},
  {Gonzalez}, {Pellegrino}, and {Terreran}]{Shrestha25}
{Shrestha}, M.; {DeSoto}, S.; {Sand}, D.J.; {Williams}, G.G.; {Hoffman}, J.L.;
  {Smith}, P.S.; {McCall}, C.; {Maund}, J.R.; {Steele}, I.A.; {Wiersema}, K.;
  et~al.
\newblock {Spectropolarimetry of SN 2023ixf Reveals Both Circumstellar Material
  and an Aspherical Helium Core}.
\newblock {\em ApJL} {\bf 2025}, {\em 982},~L32,
  \href{http://arxiv.org/abs/2410.08199}{{\normalfont
  [arXiv:astro-ph.HE/2410.08199]}}.
\newblock {\url{https://doi.org/10.3847/2041-8213/adbb63}}.

\bibitem[{Fang} et~al.(2025){Fang}, {Moriya}, {Ferrari}, {Maeda}, {Folatelli},
  {Ertini}, {Kuncarayakti}, {Andrews}, and {Matsumoto}]{Fang25}
{Fang}, Q.; {Moriya}, T.J.; {Ferrari}, L.; {Maeda}, K.; {Folatelli}, G.;
  {Ertini}, K.Y.; {Kuncarayakti}, H.; {Andrews}, J.E.; {Matsumoto}, T.
\newblock {Diversity in Hydrogen-rich Envelope Mass of Type II Supernovae. II.
  SN 2023ixf as Explosion of Partially Stripped Intermediate Massive Star}.
\newblock {\em ApJ} {\bf 2025}, {\em 978},~36,
  \href{http://arxiv.org/abs/2409.03540}{{\normalfont
  [arXiv:astro-ph.HE/2409.03540]}}.
\newblock {\url{https://doi.org/10.3847/1538-4357/ad8d5a}}.

\bibitem[{Kumar} et~al.(2025){Kumar}, {Dastidar}, {Maund}, {Singleton}, and
  {Sun}]{Kumar25}
{Kumar}, A.; {Dastidar}, R.; {Maund}, J.R.; {Singleton}, A.J.; {Sun}, N.C.
\newblock {Signatures of the shock interaction as an additional power source in
  the nebular spectra of SN 2023ixf}.
\newblock {\em MNRAS} {\bf 2025}, {\em 538},~659--670,
  \href{http://arxiv.org/abs/2412.03509}{{\normalfont
  [arXiv:astro-ph.HE/2412.03509]}}.
\newblock {\url{https://doi.org/10.1093/mnras/staf312}}.

\bibitem[{Folatelli} et~al.(2025){Folatelli}, {Ferrari}, {Ertini},
  {Kuncarayakti}, and {Maeda}]{Folatelli25}
{Folatelli}, G.; {Ferrari}, L.; {Ertini}, K.; {Kuncarayakti}, H.; {Maeda}, K.
\newblock {SN 2023ixf: interaction signatures in the spectrum at 445 days}.
\newblock {\em arXiv e-prints} {\bf 2025}, p. arXiv:2502.10534,
  \href{http://arxiv.org/abs/2502.10534}{{\normalfont
  [arXiv:astro-ph.SR/2502.10534]}}.
\newblock {\url{https://doi.org/10.48550/arXiv.2502.10534}}.

\bibitem[{Michel} et~al.(2025){Michel}, {Mazzali}, {Perley}, {Hinds}, and
  {Wise}]{Michel25}
{Michel}, P.D.; {Mazzali}, P.A.; {Perley}, D.A.; {Hinds}, K.R.; {Wise}, J.L.
\newblock {The nebular spectra of SN 2023ixf: a lower mass, partially stripped
  progenitor may be the result of binary interaction}.
\newblock {\em MNRAS} {\bf 2025}, {\em 539},~633--649,
  \href{http://arxiv.org/abs/2503.13017}{{\normalfont
  [arXiv:astro-ph.HE/2503.13017]}}.
\newblock {\url{https://doi.org/10.1093/mnras/staf443}}.

\bibitem[{Shrestha} et~al.(2024){Shrestha}, {Pearson}, {Wyatt}, {Sand},
  {Hosseinzadeh}, {Bostroem}, {Andrews}, {Dong}, {Hoang}, {Janzen}, {Jencson},
  {Lundquist}, {Mehta}, {Retamal}, {Valenti}, {Rastinejad}, {Daly}, {Porter},
  {Hinz}, {Self}, {Weiner}, {Williams}, {Hiramatsu}, {Howell}, {McCully},
  {Gonzalez}, {Pellegrino}, {Terreran}, {Newsome}, {Farah}, {Itagaki}, {Jha},
  {Kwok}, {Smith}, {Schwab}, {Rho}, and {Yang}]{Shrestha24}
{Shrestha}, M.; {Pearson}, J.; {Wyatt}, S.; {Sand}, D.J.; {Hosseinzadeh}, G.;
  {Bostroem}, K.A.; {Andrews}, J.E.; {Dong}, Y.; {Hoang}, E.; {Janzen}, D.;
  et~al.
\newblock {Evidence of Weak Circumstellar Medium Interaction in the Type II SN
  2023axu}.
\newblock {\em ApJ} {\bf 2024}, {\em 961},~247,
  \href{http://arxiv.org/abs/2310.00162}{{\normalfont
  [arXiv:astro-ph.HE/2310.00162]}}.
\newblock {\url{https://doi.org/10.3847/1538-4357/ad11e1}}.

\bibitem[{Van Dyk} et~al.(2024){Van Dyk}, {Szalai}, {Cutri}, {Kirkpatrick},
  {Grillmair}, {Fajardo-Acosta}, {Masiero}, {Mainzer}, {Gelino}, {Vink{\'o}},
  {Jo{\'o}}, {P{\'a}l}, {K{\"o}nyves-T{\'o}th}, {Kriskovics}, {Szak{\'a}ts},
  {Vida}, {Zheng}, {Brink}, and {Filippenko}]{vandyk24}
{Van Dyk}, S.D.; {Szalai}, T.; {Cutri}, R.M.; {Kirkpatrick}, J.D.; {Grillmair},
  C.J.; {Fajardo-Acosta}, S.B.; {Masiero}, J.R.; {Mainzer}, A.K.; {Gelino},
  C.R.; {Vink{\'o}}, J.;  et~al.
\newblock {NEOWISE-R Caught the Luminous SN 2023ixf in Messier 101}.
\newblock {\em ApJ} {\bf 2024}, {\em 977},~98,
  \href{http://arxiv.org/abs/2406.18005}{{\normalfont
  [arXiv:astro-ph.SR/2406.18005]}}.
\newblock {\url{https://doi.org/10.3847/1538-4357/ad8cd8}}.

\bibitem[{Hinds} et~al.(2025){Hinds}, {Perley}, {Sollerman}, {Miller},
  {Fremling}, {Moriya}, {Das}, {Qin}, {Bellm}, {Chen}, {Coughlin},
  {Jacobson-Gal\textbackslash'an}, {Kasliwal}, {Kulkarni}, {Mahabal}, {Masci},
  {Priscila}, {Pessi}, {Purdum}, {Riddle}, {Singh}, {Smith}, and
  {Sravan}]{Hinds25}
{Hinds}, K.R.; {Perley}, D.; {Sollerman}, J.; {Miller}, A.; {Fremling}, C.;
  {Moriya}, T.; {Das}, K.; {Qin}, Y.J.; {Bellm}, E.; {Chen}, X.T.;  et~al.
\newblock {Inferring CSM Properties of Type II SNe Using a Magnitude-Limited
  ZTF Sample}.
\newblock {\em arXiv e-prints} {\bf 2025}, p. arXiv:2503.19969,
  \href{http://arxiv.org/abs/2503.19969}{{\normalfont
  [arXiv:astro-ph.HE/2503.19969]}}.
\newblock {\url{https://doi.org/10.48550/arXiv.2503.19969}}.

\bibitem[{Yang} et~al.(2024){Yang}, {Liu}, {Pan}, {Er}, {Liu}, {Fang}, {Du},
  {Cai}, {Xu}, {Chen}, {Zou}, {Guo}, {Liu}, {Cheng}, {Kumar}, and
  {Liu}]{Yang24}
{Yang}, Y.P.; {Liu}, X.; {Pan}, Y.; {Er}, X.; {Liu}, D.; {Fang}, Y.; {Du}, G.;
  {Cai}, Y.; {Xu}, X.; {Chen}, X.;  et~al.
\newblock {Multiband Simultaneous Photometry of Type II SN 2023ixf with
  Mephisto and the Twin 50 cm Telescopes}.
\newblock {\em ApJ} {\bf 2024}, {\em 969},~126,
  \href{http://arxiv.org/abs/2405.08327}{{\normalfont
  [arXiv:astro-ph.HE/2405.08327]}}.
\newblock {\url{https://doi.org/10.3847/1538-4357/ad4be3}}.

\bibitem[{Hsu} et~al.(2024){Hsu}, {Smith}, {Goldberg}, {Bostroem},
  {Hosseinzadeh}, {Sand}, {Pearson}, {Hiramatsu}, {Andrews}, {Beasor}, {Dong},
  {Farah}, {Galbany}, {Gomez}, {Padilla Gonzalez}, {Guti{\'e}rrez}, {Howell},
  {K{\"o}nyves-T{\'o}th}, {McCully}, {Newsome}, {Shrestha}, {Terreran},
  {Villar}, and {Wang}]{Hsu24}
{Hsu}, B.; {Smith}, N.; {Goldberg}, J.A.; {Bostroem}, K.A.; {Hosseinzadeh}, G.;
  {Sand}, D.J.; {Pearson}, J.; {Hiramatsu}, D.; {Andrews}, J.E.; {Beasor},
  E.R.;  et~al.
\newblock {One Year of SN 2023ixf: Breaking Through the Degenerate Parameter
  Space in Light-Curve Models with Pulsating Progenitors}.
\newblock {\em arXiv e-prints} {\bf 2024}, p. arXiv:2408.07874,
  \href{http://arxiv.org/abs/2408.07874}{{\normalfont
  [arXiv:astro-ph.HE/2408.07874]}}.
\newblock {\url{https://doi.org/10.48550/arXiv.2408.07874}}.

\bibitem[{Forde} and {Goldberg}(2025)]{Forde25}
{Forde}, S.; {Goldberg}, J.A.
\newblock {Modeling Supernova 2023ixf: Lightcurve Degeneracies and
  Morphological Differences}.
\newblock {\em arXiv e-prints} {\bf 2025}, p. arXiv:2504.12421,
  \href{http://arxiv.org/abs/2504.12421}{{\normalfont
  [arXiv:astro-ph.SR/2504.12421]}}.
\newblock {\url{https://doi.org/10.48550/arXiv.2504.12421}}.

\bibitem[{Li} et~al.(2025){Li}, {Wang}, {Yang}, {Pastorello}, {Reguitti},
  {Valerin}, {Ochner}, {Cai}, {Iijima}, {Munari}, {Salmaso}, {Farina},
  {Cazzola}, {Trabacchin}, {Fiscale}, {Ciroi}, {Mura}, {Siviero}, {Cabras},
  {Pabst}, {Taubenberger}, {Vogl}, {Fiorin}, {Liu}, {Chen}, {Xiang}, {Mo},
  {Li}, {Wang}, {Zhang}, {Zhai}, {Mirzaqulov}, {Ehgamberdiev}, {Filippenko},
  {Yan}, {Hu}, {Ma}, {Xia}, {Gao}, and {Li}]{Li25}
{Li}, G.; {Wang}, X.; {Yang}, Y.; {Pastorello}, A.; {Reguitti}, A.; {Valerin},
  G.; {Ochner}, P.; {Cai}, Y.; {Iijima}, T.; {Munari}, U.;  et~al.
\newblock {Optical and Near-infrared Observations of SN 2023ixf for over 600
  days after the Explosion}.
\newblock {\em arXiv e-prints} {\bf 2025}, p. arXiv:2504.03856,
  \href{http://arxiv.org/abs/2504.03856}{{\normalfont
  [arXiv:astro-ph.HE/2504.03856]}}.
\newblock {\url{https://doi.org/10.48550/arXiv.2504.03856}}.

\bibitem[{Grefenstette} et~al.(2023){Grefenstette}, {Brightman}, {Earnshaw},
  {Harrison}, and {Margutti}]{Grefenstette23}
{Grefenstette}, B.W.; {Brightman}, M.; {Earnshaw}, H.P.; {Harrison}, F.A.;
  {Margutti}, R.
\newblock {Early Hard X-Rays from the Nearby Core-collapse Supernova SN
  2023ixf}.
\newblock {\em ApJL} {\bf 2023}, {\em 952},~L3,
  \href{http://arxiv.org/abs/2306.04827}{{\normalfont
  [arXiv:astro-ph.HE/2306.04827]}}.
\newblock {\url{https://doi.org/10.3847/2041-8213/acdf4e}}.

\bibitem[{Chandra} et~al.(2024){Chandra}, {Chevalier}, {Maeda}, {Ray}, and
  {Nayana}]{Chandra24}
{Chandra}, P.; {Chevalier}, R.A.; {Maeda}, K.; {Ray}, A.K.; {Nayana}, A.J.
\newblock {Chandra's Insights into SN 2023ixf}.
\newblock {\em ApJL} {\bf 2024}, {\em 963},~L4,
  \href{http://arxiv.org/abs/2311.04384}{{\normalfont
  [arXiv:astro-ph.HE/2311.04384]}}.
\newblock {\url{https://doi.org/10.3847/2041-8213/ad275d}}.

\bibitem[{Panjkov} et~al.(2024){Panjkov}, {Auchettl}, {Shappee}, {Do}, {Lopez},
  and {Beacom}]{Panjkov24}
{Panjkov}, S.; {Auchettl}, K.; {Shappee}, B.J.; {Do}, A.; {Lopez}, L.;
  {Beacom}, J.F.
\newblock {Probing the soft X-ray properties and multi-wavelength variability
  of SN2023ixf and its progenitor}.
\newblock {\em PASA} {\bf 2024}, {\em 41},~e059,
  \href{http://arxiv.org/abs/2308.13101}{{\normalfont
  [arXiv:astro-ph.HE/2308.13101]}}.
\newblock {\url{https://doi.org/10.1017/pasa.2024.66}}.

\bibitem[{A.~J.} et~al.(2025){A.~J.}, {Margutti}, {Wiston}, {Chornock},
  {Campana}, {Laskar}, {Murase}, {Krips}, {Migliori}, {Tsuna}, {Alexander},
  {Chandra}, {Bietenholz}, {Berger}, {Chevalier}, {De Colle}, {Dessart},
  {Diesing}, {Grefenstette}, {Jacobson-Gal{\'a}n}, {Maeda}, {Marcote},
  {Matthews}, {Milisavljevic}, {Ray}, {Reguitti}, and {Polzin}]{Nayana25}
{A.~J.}, N.; {Margutti}, R.; {Wiston}, E.; {Chornock}, R.; {Campana}, S.;
  {Laskar}, T.; {Murase}, K.; {Krips}, M.; {Migliori}, G.; {Tsuna}, D.;  et~al.
\newblock {Dinosaur in a Haystack: X-Ray View of the Entrails of SN 2023ixf and
  the Radio Afterglow of Its Interaction with the Medium Spawned by the
  Progenitor Star (Paper I)}.
\newblock {\em ApJ} {\bf 2025}, {\em 985},~51.
\newblock {\url{https://doi.org/10.3847/1538-4357/adc2fb}}.

\bibitem[{Berger} et~al.(2023){Berger}, {Keating}, {Margutti}, {Maeda},
  {Alexander}, {Cendes}, {Eftekhari}, {Gurwell}, {Hiramatsu}, {Ho}, {Laskar},
  {Rao}, and {Williams}]{Berger23}
{Berger}, E.; {Keating}, G.K.; {Margutti}, R.; {Maeda}, K.; {Alexander}, K.D.;
  {Cendes}, Y.; {Eftekhari}, T.; {Gurwell}, M.; {Hiramatsu}, D.; {Ho}, A.Y.Q.;
  et~al.
\newblock {Millimeter Observations of the Type II SN 2023ixf: Constraints on
  the Proximate Circumstellar Medium}.
\newblock {\em ApJL} {\bf 2023}, {\em 951},~L31,
  \href{http://arxiv.org/abs/2306.09311}{{\normalfont
  [arXiv:astro-ph.HE/2306.09311]}}.
\newblock {\url{https://doi.org/10.3847/2041-8213/ace0c4}}.

\bibitem[{Matthews} et~al.(2023){Matthews}, {Margutti}, {Alexander}, {Bright},
  {Cendes}, {Berger}, {Lasker}, {Drout}, and {Milisavljevic}]{Matthews23}
{Matthews}, D.; {Margutti}, R.; {Alexander}, K.D.; {Bright}, J.; {Cendes}, Y.;
  {Berger}, E.; {Lasker}, T.; {Drout}, M.; {Milisavljevic}, D.
\newblock {VLA 10 GHz Observations of SN2023ixf}.
\newblock {\em Transient Name Server AstroNote} {\bf 2023}, {\em 146},~1.

\bibitem[{Timmerman} et~al.(2024){Timmerman}, {Arias}, and
  {Botteon}]{Timmerman24}
{Timmerman}, R.; {Arias}, M.; {Botteon}, A.
\newblock {LOFAR Non-detections of SN 2023ixf in its First Year
  Post-explosion}.
\newblock {\em Research Notes of the American Astronomical Society} {\bf 2024},
  {\em 8},~311,  \href{http://arxiv.org/abs/2412.14275}{{\normalfont
  [arXiv:astro-ph.HE/2412.14275]}}.
\newblock {\url{https://doi.org/10.3847/2515-5172/ad9eae}}.

\bibitem[{Iwata} et~al.(2025){Iwata}, {Akimoto}, {Matsuoka}, {Maeda},
  {Yonekura}, {Tominaga}, {Moriya}, {Fujisawa}, {Niinuma}, {Yoon}, {Lee},
  {Jung}, and {Byun}]{Iwata25}
{Iwata}, Y.; {Akimoto}, M.; {Matsuoka}, T.; {Maeda}, K.; {Yonekura}, Y.;
  {Tominaga}, N.; {Moriya}, T.J.; {Fujisawa}, K.; {Niinuma}, K.; {Yoon}, S.C.;
  et~al.
\newblock {Radio Follow-up Observations of SN 2023ixf by Japanese and Korean
  Very Long Baseline Interferometers}.
\newblock {\em ApJ} {\bf 2025}, {\em 978},~138,
  \href{http://arxiv.org/abs/2411.07542}{{\normalfont
  [arXiv:astro-ph.HE/2411.07542]}}.
\newblock {\url{https://doi.org/10.3847/1538-4357/ad9a62}}.

\bibitem[{Lee} et~al.(2024){Lee}, {Lee}, {Paragi}, {Orosz}, {Oh}, and
  {Kim}]{Lee24}
{Lee}, D.; {Lee}, S.Y.; {Paragi}, Z.; {Orosz}, G.; {Oh}, J.; {Kim}, J.Y.
\newblock {EVN 5 GHz e-VLBI Observations of SN2023ixf in M101}.
\newblock {\em Research Notes of the American Astronomical Society} {\bf 2024},
  {\em 8},~121.
\newblock {\url{https://doi.org/10.3847/2515-5172/ad454e}}.

\bibitem[{Mart{\'\i}-Devesa} et~al.(2024){Mart{\'\i}-Devesa}, {Cheung}, {Di
  Lalla}, {Renaud}, {Principe}, {Omodei}, and {Acero}]{Marti24}
{Mart{\'\i}-Devesa}, G.; {Cheung}, C.C.; {Di Lalla}, N.; {Renaud}, M.;
  {Principe}, G.; {Omodei}, N.; {Acero}, F.
\newblock {Early-time {\ensuremath{\gamma}}-ray constraints on cosmic-ray
  acceleration in the core-collapse SN 2023ixf with the Fermi Large Area
  Telescope}.
\newblock {\em A\&A} {\bf 2024}, {\em 686},~A254,
  \href{http://arxiv.org/abs/2404.10487}{{\normalfont
  [arXiv:astro-ph.HE/2404.10487]}}.
\newblock {\url{https://doi.org/10.1051/0004-6361/202349061}}.

\bibitem[{Guetta} et~al.(2023){Guetta}, {Langella}, {Gagliardini}, and {Della
  Valle}]{Guetta23}
{Guetta}, D.; {Langella}, A.; {Gagliardini}, S.; {Della Valle}, M.
\newblock {Low- and High-energy Neutrinos from SN 2023ixf in M101}.
\newblock {\em ApJL} {\bf 2023}, {\em 955},~L9,
  \href{http://arxiv.org/abs/2306.14717}{{\normalfont
  [arXiv:astro-ph.HE/2306.14717]}}.
\newblock {\url{https://doi.org/10.3847/2041-8213/acf573}}.

\bibitem[{Sarmah}(2024)]{Sarmah24}
{Sarmah}, P.
\newblock {New constraints on the gamma-ray and high energy neutrino fluxes
  from the circumstellar interaction of SN 2023ixf}.
\newblock {\em JCAP} {\bf 2024}, {\em 2024},~083,
  \href{http://arxiv.org/abs/2307.08744}{{\normalfont
  [arXiv:astro-ph.HE/2307.08744]}}.
\newblock {\url{https://doi.org/10.1088/1475-7516/2024/04/083}}.

\bibitem[{Kimura} and {Moriya}(2025)]{Kimura25}
{Kimura}, S.S.; {Moriya}, T.J.
\newblock {High-energy Gamma-Ray and Neutrino Emissions from Interacting
  Supernovae Based on Radiation Hydrodynamic Simulations: A Case of SN
  2023ixf}.
\newblock {\em ApJ} {\bf 2025}, {\em 984},~103,
  \href{http://arxiv.org/abs/2409.18935}{{\normalfont
  [arXiv:astro-ph.HE/2409.18935]}}.
\newblock {\url{https://doi.org/10.3847/1538-4357/adc716}}.

\bibitem[{Ravensburg} et~al.(2024){Ravensburg}, {Carenza}, {Eckner}, and
  {Goobar}]{Ravensburg24}
{Ravensburg}, E.; {Carenza}, P.; {Eckner}, C.; {Goobar}, A.
\newblock {Constraining MeV-scale axionlike particles with Fermi-LAT
  observations of SN 2023ixf}.
\newblock {\em PRD} {\bf 2024}, {\em 109},~023018,
  \href{http://arxiv.org/abs/2306.16397}{{\normalfont
  [arXiv:astro-ph.HE/2306.16397]}}.
\newblock {\url{https://doi.org/10.1103/PhysRevD.109.023018}}.

\bibitem[{Cosentino} et~al.(2025){Cosentino}, {Pumo}, and
  {Cherubini}]{Cosentino25}
{Cosentino}, S.P.; {Pumo}, M.L.; {Cherubini}, S.
\newblock {High-Energy Neutrinos by Hydrogen-rich Supernovae interacting with
  low-massive Circumstellar Medium: The Case of SN 2023ixf}.
\newblock {\em MNRAS} {\bf 2025}.
\newblock {\url{https://doi.org/10.1093/mnras/staf861}}.

\bibitem[{Abac} et~al.(2025){Abac}, {Abbott}, {Abouelfettouh}, {Acernese},
  {Ackley}, {Adhicary}, {Adhikari}, {Adhikari}, {Adkins}, {Agarwal}, {Agathos},
  {Aghaei Abchouyeh}, {Aguiar}, {Aguilar}, {Aiello}, {Ain}, {Akutsu},
  {Albanesi}, {Alfaidi}, {Al-Jodah}, {All{\'e}n{\'e}}, {Allocca},
  {Al-Shammari}, {Altin}, {Alvarez-Lopez}, {Amato}, {Amez-Droz}, {Amorosi},
  {Amra}, {Ananyeva}, {Anderson}, {Anderson}, {Andia}, {Ando}, {Andrade},
  {Andres}, {Andr{\'e}s-Carcasona}, {Andri{\'c}}, {Anglin}, {Ansoldi},
  {Antelis}, {Antier}, {Aoumi}, {Appavuravther}, {Appert}, {Apple}, {Arai},
  {Araya}, {Araya}, {Areeda}, {Argianas}, {Aritomi}, {Armato}, {Arnaud},
  {Arogeti}, {Aronson}, {Ashton}, {Aso}, {Assiduo}, {Assis de Souza Melo},
  {Aston}, {Astone}, {Attadio}, {Aubin}, {Aultoneal}, {Avallone}, {Babak},
  {Badaracco}, {Badger}, {Bae}, {Bagnasco}, {Bagui}, {Baier}, {Baiotti},
  {Bajpai}, {Baka}, {Ball}, {Ballardin}, {Ballmer}, {Banagiri}, {Banerjee},
  {Bankar}, {Baral}, {Barayoga}, {Barish}, {Barker}, {Barneo}, {Barone},
  {Barr}, {Barsotti}, {Barsuglia}, {Barta}, {Bartoletti}, {Barton}, {Bartos},
  {Basak}, {Basalaev}, {Bassiri}, {Basti}, {Bates}, {Bawaj}, {Baxi}, {Bayley},
  {Baylor}, {Baynard}, {Bazzan}, {Bedakihale}, {Beirnaert}, {Bejger},
  {Belardinelli}, {Bell}, {Benedetto}, {Benoit}, {Bentley}, {Ben Yaala},
  {Bera}, {Berbel}, {Bergamin}, {Berger}, {Bernuzzi}, {Beroiz}, {Bersanetti},
  {Bertolini}, {Betzwieser}, {Beveridge}, {Bevins}, {Bhandare}, {Bhardwaj},
  {Bhatt}, {Bhattacharjee}, {Bhaumik}, {Bhowmick}, {Bianchi}, {Bilenko},
  {Billingsley}, {Binetti}, {Bini}, {Birnholtz}, {Biscoveanu}, {Bisht},
  {Bitossi}, {Bizouard}, {Blackburn}, {Blagg}, {Blair}, {Blair}, {Bobba},
  {Bode}, {Boileau}, {Boldrini}, {Bolingbroke}, {Bolliand}, {Bonavena},
  {Bondarescu}, {Bondu}, {Bonilla}, {Bonilla}, {Bonino}, {Bonnand}, {Booker},
  {Borchers}, {Boschi}, {Bose}, {Bossilkov}, {Boudart}, {Boudon}, {Bozzi},
  {Bradaschia}, {Brady}, {Braglia}, {Branch}, {Branchesi}, {Brandt}, {Braun},
  {Breschi}, {Briant}, {Brillet}, {Brinkmann}, {Brockill}, {Brockmueller},
  {Brooks}, {Brown}, {Brown}, {Brozzetti}, {Brunett}, {Bruno}, {Bruntz},
  {Bryant}, {Bucci}, {Buchanan}, {Bulashenko}, {Bulik}, {Bulten}, {Buonanno},
  {Burtnyk}, {Buscicchio}, {Buskulic}, {Buy}, {Byer}, and {Cabourn
  Davies}]{Abac25}
{Abac}, A.G.; {Abbott}, R.; {Abouelfettouh}, I.; {Acernese}, F.; {Ackley}, K.;
  {Adhicary}, S.; {Adhikari}, N.; {Adhikari}, R.X.; {Adkins}, V.K.; {Agarwal},
  D.;  et~al.
\newblock {Search for Gravitational Waves Emitted from SN 2023ixf}.
\newblock {\em ApJ} {\bf 2025}, {\em 985},~183,
  \href{http://arxiv.org/abs/2410.16565}{{\normalfont
  [arXiv:astro-ph.HE/2410.16565]}}.
\newblock {\url{https://doi.org/10.3847/1538-4357/adc681}}.

\bibitem[{Kilpatrick} et~al.(2023){Kilpatrick}, {Foley}, {Jacobson-Gal{\'a}n},
  {Piro}, {Smartt}, {Drout}, {Gagliano}, {Gall}, {Hjorth}, {Jones}, {Mandel},
  {Margutti}, {Ramirez-Ruiz}, {Ransome}, {Villar}, {Coulter}, {Gao},
  {Matthews}, {Taggart}, and {Zenati}]{Kilpatrick23}
{Kilpatrick}, C.D.; {Foley}, R.J.; {Jacobson-Gal{\'a}n}, W.V.; {Piro}, A.L.;
  {Smartt}, S.J.; {Drout}, M.R.; {Gagliano}, A.; {Gall}, C.; {Hjorth}, J.;
  {Jones}, D.O.;  et~al.
\newblock {SN 2023ixf in Messier 101: A Variable Red Supergiant as the
  Progenitor Candidate to a Type II Supernova}.
\newblock {\em ApJL} {\bf 2023}, {\em 952},~L23,
  \href{http://arxiv.org/abs/2306.04722}{{\normalfont
  [arXiv:astro-ph.SR/2306.04722]}}.
\newblock {\url{https://doi.org/10.3847/2041-8213/ace4ca}}.

\bibitem[{Jencson} et~al.(2023){Jencson}, {Pearson}, {Beasor}, {Lau},
  {Andrews}, {Bostroem}, {Dong}, {Engesser}, {Gomez}, {Guolo}, {Hoang},
  {Hosseinzadeh}, {Jha}, {Karambelkar}, {Kasliwal}, {Lundquist}, {Meza
  Retamal}, {Rest}, {Sand}, {Shahbandeh}, {Shrestha}, {Smith}, {Strader},
  {Valenti}, {Wang}, and {Zenati}]{Jencson23}
{Jencson}, J.E.; {Pearson}, J.; {Beasor}, E.R.; {Lau}, R.M.; {Andrews}, J.E.;
  {Bostroem}, K.A.; {Dong}, Y.; {Engesser}, M.; {Gomez}, S.; {Guolo}, M.;
  et~al.
\newblock {A Luminous Red Supergiant and Dusty Long-period Variable Progenitor
  for SN 2023ixf}.
\newblock {\em ApJL} {\bf 2023}, {\em 952},~L30,
  \href{http://arxiv.org/abs/2306.08678}{{\normalfont
  [arXiv:astro-ph.SR/2306.08678]}}.
\newblock {\url{https://doi.org/10.3847/2041-8213/ace618}}.

\bibitem[{Niu} et~al.(2023){Niu}, {Sun}, {Maund}, {Zhang}, {Zhao}, and
  {Liu}]{Niu23}
{Niu}, Z.; {Sun}, N.C.; {Maund}, J.R.; {Zhang}, Y.; {Zhao}, R.; {Liu}, J.
\newblock {The Dusty Red Supergiant Progenitor and the Local Environment of the
  Type II SN 2023ixf in M101}.
\newblock {\em ApJL} {\bf 2023}, {\em 955},~L15,
  \href{http://arxiv.org/abs/2308.04677}{{\normalfont
  [arXiv:astro-ph.SR/2308.04677]}}.
\newblock {\url{https://doi.org/10.3847/2041-8213/acf4e3}}.

\bibitem[{Soraisam} et~al.(2023){Soraisam}, {Szalai}, {Van Dyk}, {Andrews},
  {Srinivasan}, {Chun}, {Matheson}, {Scicluna}, and
  {Vasquez-Torres}]{Soraisam23}
{Soraisam}, M.D.; {Szalai}, T.; {Van Dyk}, S.D.; {Andrews}, J.E.; {Srinivasan},
  S.; {Chun}, S.H.; {Matheson}, T.; {Scicluna}, P.; {Vasquez-Torres}, D.A.
\newblock {The SN 2023ixf Progenitor in M101. I. Infrared Variability}.
\newblock {\em ApJ} {\bf 2023}, {\em 957},~64,
  \href{http://arxiv.org/abs/2306.10783}{{\normalfont
  [arXiv:astro-ph.SR/2306.10783]}}.
\newblock {\url{https://doi.org/10.3847/1538-4357/acef22}}.

\bibitem[{Van Dyk} et~al.(2024){Van Dyk}, {Srinivasan}, {Andrews}, {Soraisam},
  {Szalai}, {Howell}, {Isaacson}, {Matheson}, {Petigura}, {Scicluna},
  {Stephens}, {Van Zandt}, {Zheng}, {Chun}, and {Fillippenko}]{VanDyk23}
{Van Dyk}, S.D.; {Srinivasan}, S.; {Andrews}, J.E.; {Soraisam}, M.; {Szalai},
  T.; {Howell}, S.B.; {Isaacson}, H.; {Matheson}, T.; {Petigura}, E.;
  {Scicluna}, P.;  et~al.
\newblock {The SN 2023ixf Progenitor in M101. II. Properties}.
\newblock {\em ApJ} {\bf 2024}, {\em 968},~27,
  \href{http://arxiv.org/abs/2308.14844}{{\normalfont
  [arXiv:astro-ph.SR/2308.14844]}}.
\newblock {\url{https://doi.org/10.3847/1538-4357/ad414b}}.

\bibitem[{Qin} et~al.(2024){Qin}, {Zhang}, {Bloom}, {Sollerman}, {Zimmerman},
  {Irani}, {Schulze}, {Gal-Yam}, {Kasliwal}, {Coughlin}, {Perley}, {Fremling},
  and {Kulkarni}]{Qin23}
{Qin}, Y.J.; {Zhang}, K.; {Bloom}, J.; {Sollerman}, J.; {Zimmerman}, E.A.;
  {Irani}, I.; {Schulze}, S.; {Gal-Yam}, A.; {Kasliwal}, M.; {Coughlin}, M.W.;
  et~al.
\newblock {The progenitor star of SN 2023ixf: a massive red supergiant with
  enhanced, episodic pre-supernova mass loss}.
\newblock {\em MNRAS} {\bf 2024}, {\em 534},~271--280,
  \href{http://arxiv.org/abs/2309.10022}{{\normalfont
  [arXiv:astro-ph.SR/2309.10022]}}.
\newblock {\url{https://doi.org/10.1093/mnras/stae2012}}.

\bibitem[{Pledger} and {Shara}(2023)]{Pledger23}
{Pledger}, J.L.; {Shara}, M.M.
\newblock {Possible Detection of the Progenitor of the Type II Supernova SN
  2023ixf}.
\newblock {\em ApJL} {\bf 2023}, {\em 953},~L14,
  \href{http://arxiv.org/abs/2305.14447}{{\normalfont
  [arXiv:astro-ph.SR/2305.14447]}}.
\newblock {\url{https://doi.org/10.3847/2041-8213/ace88b}}.

\bibitem[{Xiang} et~al.(2024){Xiang}, {Mo}, {Wang}, {Wang}, {Zhang}, {Lin}, and
  {Wang}]{Xiang24}
{Xiang}, D.; {Mo}, J.; {Wang}, L.; {Wang}, X.; {Zhang}, J.; {Lin}, H.; {Wang},
  L.
\newblock {The dusty and extremely red progenitor of the type II supernova
  2023ixf in Messier 101}.
\newblock {\em Science China Physics, Mechanics, and Astronomy} {\bf 2024},
  {\em 67},~219514,  \href{http://arxiv.org/abs/2309.01389}{{\normalfont
  [arXiv:astro-ph.SR/2309.01389]}}.
\newblock {\url{https://doi.org/10.1007/s11433-023-2267-0}}.

\bibitem[{Ransome} et~al.(2024){Ransome}, {Villar}, {Tartaglia}, {Gonzalez},
  {Jacobson-Gal{\'a}n}, {Kilpatrick}, {Margutti}, {Foley}, {Grayling}, {Ni},
  {Yarza}, {Ye}, {Auchettl}, {de Boer}, {Chambers}, {Coulter}, {Drout},
  {Farias}, {Gall}, {Gao}, {Huber}, {Ibik}, {Jones}, {Khetan}, {Lin},
  {Politsch}, {Raimundo}, {Rest}, {Wainscoat}, {Yadavalli}, and
  {Zenati}]{Ransome24}
{Ransome}, C.L.; {Villar}, V.A.; {Tartaglia}, A.; {Gonzalez}, S.J.;
  {Jacobson-Gal{\'a}n}, W.V.; {Kilpatrick}, C.D.; {Margutti}, R.; {Foley},
  R.J.; {Grayling}, M.; {Ni}, Y.Q.;  et~al.
\newblock {SN 2023ixf in Messier 101: The Twilight Years of the Progenitor as
  Seen by Pan-STARRS}.
\newblock {\em ApJ} {\bf 2024}, {\em 965},~93,
  \href{http://arxiv.org/abs/2312.04426}{{\normalfont
  [arXiv:astro-ph.SR/2312.04426]}}.
\newblock {\url{https://doi.org/10.3847/1538-4357/ad2df7}}.

\bibitem[{Jacobson-Gal{\'a}n} et~al.(2022){Jacobson-Gal{\'a}n}, {Dessart},
  {Jones}, {Margutti}, {Coppejans}, {Dimitriadis}, {Foley}, {Kilpatrick},
  {Matthews}, {Rest}, {Terreran}, {Aleo}, {Auchettl}, {Blanchard}, {Coulter},
  {Davis}, {de Boer}, {DeMarchi}, {Drout}, {Earl}, {Gagliano}, {Gall},
  {Hjorth}, {Huber}, {Ibik}, {Milisavljevic}, {Pan}, {Rest}, {Ridden-Harper},
  {Rojas-Bravo}, {Siebert}, {Smith}, {Taggart}, {Tinyanont}, {Wang}, and
  {Zenati}]{wjg22}
{Jacobson-Gal{\'a}n}, W.V.; {Dessart}, L.; {Jones}, D.O.; {Margutti}, R.;
  {Coppejans}, D.L.; {Dimitriadis}, G.; {Foley}, R.J.; {Kilpatrick}, C.D.;
  {Matthews}, D.J.; {Rest}, S.;  et~al.
\newblock {Final Moments. I. Precursor Emission, Envelope Inflation, and
  Enhanced Mass Loss Preceding the Luminous Type II Supernova 2020tlf}.
\newblock {\em ApJ} {\bf 2022}, {\em 924},~15,
  \href{http://arxiv.org/abs/2109.12136}{{\normalfont
  [arXiv:astro-ph.HE/2109.12136]}}.
\newblock {\url{https://doi.org/10.3847/1538-4357/ac3f3a}}.

\bibitem[{Strotjohann} et~al.(2021){Strotjohann}, {Ofek}, {Gal-Yam}, {Bruch},
  {Schulze}, {Shaviv}, {Sollerman}, {Filippenko}, {Yaron}, {Fremling},
  {Nordin}, {Kool}, {Perley}, {Ho}, {Yang}, {Yao}, {Soumagnac}, {Graham},
  {Barbarino}, {Tartaglia}, {De}, {Goldstein}, {Cook}, {Brink}, {Taggart},
  {Yan}, {Lunnan}, {Kasliwal}, {Kulkarni}, {Nugent}, {Masci}, {Rosnet},
  {Adams}, {Andreoni}, {Bagdasaryan}, {Bellm}, {Burdge}, {Duev}, {Dugas},
  {Frederick}, {Goldwasser}, {Hankins}, {Irani}, {Karambelkar}, {Kupfer},
  {Liang}, {Neill}, {Porter}, {Riddle}, {Sharma}, {Short}, {Taddia},
  {Tzanidakis}, {van Roestel}, {Walters}, and {Zhuang}]{Strotjohann21}
{Strotjohann}, N.L.; {Ofek}, E.O.; {Gal-Yam}, A.; {Bruch}, R.; {Schulze}, S.;
  {Shaviv}, N.; {Sollerman}, J.; {Filippenko}, A.V.; {Yaron}, O.; {Fremling},
  C.;  et~al.
\newblock {Bright, Months-long Stellar Outbursts Announce the Explosion of
  Interaction-powered Supernovae}.
\newblock {\em ApJ} {\bf 2021}, {\em 907},~99,
  \href{http://arxiv.org/abs/2010.11196}{{\normalfont
  [arXiv:astro-ph.HE/2010.11196]}}.
\newblock {\url{https://doi.org/10.3847/1538-4357/abd032}}.

\bibitem[{Neustadt} et~al.(2024){Neustadt}, {Kochanek}, and
  {Smith}]{Neustadt24}
{Neustadt}, J.M.M.; {Kochanek}, C.S.; {Smith}, M.R.
\newblock {Constraints on pre-SN outbursts from the progenitor of SN 2023ixf
  using the large binocular telescope}.
\newblock {\em MNRAS} {\bf 2024}, {\em 527},~5366--5373,
  \href{http://arxiv.org/abs/2306.06162}{{\normalfont
  [arXiv:astro-ph.HE/2306.06162]}}.
\newblock {\url{https://doi.org/10.1093/mnras/stad3073}}.

\bibitem[{Dong} et~al.(2023){Dong}, {Sand}, {Valenti}, {Bostroem}, {Andrews},
  {Hosseinzadeh}, {Hoang}, {Janzen}, {Jencson}, {Lundquist}, {Meza Retamal},
  {Pearson}, {Shrestha}, {Haislip}, {Kouprianov}, and {Reichart}]{Dong23}
{Dong}, Y.; {Sand}, D.J.; {Valenti}, S.; {Bostroem}, K.A.; {Andrews}, J.E.;
  {Hosseinzadeh}, G.; {Hoang}, E.; {Janzen}, D.; {Jencson}, J.E.; {Lundquist},
  M.;  et~al.
\newblock {A Comprehensive Optical Search for Pre-explosion Outbursts from the
  Quiescent Progenitor of SN 2023ixf}.
\newblock {\em ApJ} {\bf 2023}, {\em 957},~28,
  \href{http://arxiv.org/abs/2307.02539}{{\normalfont
  [arXiv:astro-ph.HE/2307.02539]}}.
\newblock {\url{https://doi.org/10.3847/1538-4357/acef18}}.

\bibitem[{Rest} et~al.(2025){Rest}, {Rest}, {Kilpatrick}, {Jencson}, {von
  Coelln}, {Strolger}, {Smartt}, {Anderson}, {Clocchiatti}, {Coulter},
  {Denneau}, {Gomez}, {Heinze}, {Ridden-Harper}, {Smith}, {Stalder}, {Tonry},
  {Wang}, and {Zenati}]{Rest25}
{Rest}, S.; {Rest}, A.; {Kilpatrick}, C.D.; {Jencson}, J.E.; {von Coelln}, S.;
  {Strolger}, L.; {Smartt}, S.; {Anderson}, J.P.; {Clocchiatti}, A.; {Coulter},
  D.A.;  et~al.
\newblock {ATClean: A Novel Method for Detecting Low-luminosity Transients and
  Application to Pre-explosion Counterparts from SN 2023ixf}.
\newblock {\em ApJ} {\bf 2025}, {\em 979},~114,
  \href{http://arxiv.org/abs/2405.03747}{{\normalfont
  [arXiv:astro-ph.HE/2405.03747]}}.
\newblock {\url{https://doi.org/10.3847/1538-4357/ad973d}}.

\bibitem[{Flinner} et~al.(2023){Flinner}, {Tucker}, {Beacom}, and
  {Shappee}]{Flinner23}
{Flinner}, N.; {Tucker}, M.A.; {Beacom}, J.F.; {Shappee}, B.J.
\newblock {No UV-bright Eruptions from SN 2023ixf in GALEX Imaging 15-20 yr
  Before Explosion}.
\newblock {\em Research Notes of the American Astronomical Society} {\bf 2023},
  {\em 7},~174,  \href{http://arxiv.org/abs/2308.08403}{{\normalfont
  [arXiv:astro-ph.HE/2308.08403]}}.
\newblock {\url{https://doi.org/10.3847/2515-5172/acefc4}}.

\bibitem[{Bersten} et~al.(2024){Bersten}, {Orellana}, {Folatelli}, {Martinez},
  {Piccirilli}, {Regna}, {Rom{\'a}n Aguilar}, and {Ertini}]{Bersten24}
{Bersten}, M.C.; {Orellana}, M.; {Folatelli}, G.; {Martinez}, L.; {Piccirilli},
  M.P.; {Regna}, T.; {Rom{\'a}n Aguilar}, L.M.; {Ertini}, K.
\newblock {The progenitor of SN 2023ixf from hydrodynamical modeling}.
\newblock {\em A\&A} {\bf 2024}, {\em 681},~L18,
  \href{http://arxiv.org/abs/2310.14407}{{\normalfont
  [arXiv:astro-ph.SR/2310.14407]}}.
\newblock {\url{https://doi.org/10.1051/0004-6361/202348183}}.

\bibitem[{Moriya} and {Singh}(2024)]{Moriya24}
{Moriya}, T.J.; {Singh}, A.
\newblock {Progenitor and explosion properties of SN 2023ixf estimated based on
  a light-curve model grid of Type II supernovae}.
\newblock {\em PASJ} {\bf 2024}, {\em 76},~1050--1058,
  \href{http://arxiv.org/abs/2406.00928}{{\normalfont
  [arXiv:astro-ph.HE/2406.00928]}}.
\newblock {\url{https://doi.org/10.1093/pasj/psae070}}.

\bibitem[{Jerkstrand} et~al.(2014){Jerkstrand}, {Smartt}, {Fraser}, {Fransson},
  {Sollerman}, {Taddia}, and {Kotak}]{Jerkstrand14}
{Jerkstrand}, A.; {Smartt}, S.J.; {Fraser}, M.; {Fransson}, C.; {Sollerman},
  J.; {Taddia}, F.; {Kotak}, R.
\newblock {The nebular spectra of SN 2012aw and constraints on stellar
  nucleosynthesis from oxygen emission lines}.
\newblock {\em MNRAS} {\bf 2014}, {\em 439},~3694--3703,
  \href{http://arxiv.org/abs/1311.2031}{{\normalfont
  [arXiv:astro-ph.SR/1311.2031]}}.
\newblock {\url{https://doi.org/10.1093/mnras/stu221}}.

\bibitem[{Dessart} et~al.(2021){Dessart}, {Hillier}, {Sukhbold}, {Woosley}, and
  {Janka}]{Dessart21}
{Dessart}, L.; {Hillier}, D.J.; {Sukhbold}, T.; {Woosley}, S.; {Janka}, H.T.
\newblock {The explosion of 9$-$29$M_\odot$ stars as Type II supernovae :
  results from radiative-transfer modeling at one year after explosion}.
\newblock {\em arXiv e-prints} {\bf 2021}, p. arXiv:2105.13029,
  \href{http://arxiv.org/abs/2105.13029}{{\normalfont
  [arXiv:astro-ph.SR/2105.13029]}}.

\bibitem[{Ferrari} et~al.(2024){Ferrari}, {Folatelli}, {Ertini},
  {Kuncarayakti}, and {Andrews}]{Ferrari24}
{Ferrari}, L.; {Folatelli}, G.; {Ertini}, K.; {Kuncarayakti}, H.; {Andrews},
  J.E.
\newblock {Progenitor mass and ejecta asymmetry of supernova 2023ixf from
  nebular spectroscopy}.
\newblock {\em A\&A} {\bf 2024}, {\em 687},~L20,
  \href{http://arxiv.org/abs/2406.00130}{{\normalfont
  [arXiv:astro-ph.SR/2406.00130]}}.
\newblock {\url{https://doi.org/10.1051/0004-6361/202450440}}.

\bibitem[{Jacobson-Gal{\'a}n} et~al.(2025){Jacobson-Gal{\'a}n}, {Dessart},
  {Davis}, {Bostroem}, {Kilpatrick}, {Margutti}, {Filippenko}, {Foley},
  {Chornock}, {Terreran}, {Hiramatsu}, {Newsome}, {Padilla Gonzalez},
  {Pellegrino}, {Howell}, {Anderson}, {Angus}, {Auchettl}, {Brink}, {Cartier},
  {Coulter}, {de Boer}, {Drout}, {Earl}, {Ertini}, {Farah}, {Farias}, {Gall},
  {Gao}, {Gerlach}, {Guo}, {Haynie}, {Hosseinzadeh}, {Ibik}, {Jha}, {Jones},
  {Langeroodi}, {LeBaron}, {Magnier}, {Piro}, {Raimundo}, {Rest}, {Rest},
  {Rich}, {Rojas-Bravo}, {Sears}, {Taggart}, {Villar}, {Wainscoat}, {Wang},
  {Wasserman}, {Yan}, {Yang}, {Zhang}, and {Zheng}]{wjg25a}
{Jacobson-Gal{\'a}n}, W.V.; {Dessart}, L.; {Davis}, K.W.; {Bostroem}, K.A.;
  {Kilpatrick}, C.D.; {Margutti}, R.; {Filippenko}, A.V.; {Foley}, R.J.;
  {Chornock}, R.; {Terreran}, G.;  et~al.
\newblock {Final Moments III: Explosion Properties and Progenitor Constraints
  of CSM-Interacting Type II Supernovae}.
\newblock {\em arXiv e-prints} {\bf 2025}, p. arXiv:2505.04698,
  \href{http://arxiv.org/abs/2505.04698}{{\normalfont
  [arXiv:astro-ph.HE/2505.04698]}}.
\newblock {\url{https://doi.org/10.48550/arXiv.2505.04698}}.

\bibitem[{Martinez} et~al.(2024){Martinez}, {Bersten}, {Folatelli}, {Orellana},
  and {Ertini}]{Martinez24}
{Martinez}, L.; {Bersten}, M.C.; {Folatelli}, G.; {Orellana}, M.; {Ertini}, K.
\newblock {Circumstellar interaction models for the early bolometric light
  curve of SN 2023ixf}.
\newblock {\em A\&A} {\bf 2024}, {\em 683},~A154,
  \href{http://arxiv.org/abs/2310.08733}{{\normalfont
  [arXiv:astro-ph.SR/2310.08733]}}.
\newblock {\url{https://doi.org/10.1051/0004-6361/202348142}}.

\bibitem[{Hu} et~al.(2025){Hu}, {Wang}, and {Wang}]{Hu25}
{Hu}, M.; {Wang}, L.; {Wang}, X.
\newblock {A Shock Crashing into Confined Dense Circumstellar Matter Brightens
  the Nascent SN 2023ixf}.
\newblock {\em ApJ} {\bf 2025}, {\em 984},~44,
  \href{http://arxiv.org/abs/2411.06351}{{\normalfont
  [arXiv:astro-ph.HE/2411.06351]}}.
\newblock {\url{https://doi.org/10.3847/1538-4357/adc802}}.

\bibitem[{Hiramatsu} et~al.(2023){Hiramatsu}, {Tsuna}, {Berger}, {Itagaki},
  {Goldberg}, {Gomez}, {Kishalay}, {Hosseinzadeh}, {Bostroem}, {Brown},
  {Arcavi}, {Bieryla}, {Blanchard}, {Esquerdo}, {Farah}, {Howell}, {Matsumoto},
  {McCully}, {Newsome}, {Gonzalez}, {Pellegrino}, {Rhee}, {Terreran},
  {Vink{\'o}}, and {Wheeler}]{Hiramatsu23}
{Hiramatsu}, D.; {Tsuna}, D.; {Berger}, E.; {Itagaki}, K.; {Goldberg}, J.A.;
  {Gomez}, S.; {Kishalay}, D.; {Hosseinzadeh}, G.; {Bostroem}, K.A.; {Brown},
  P.J.;  et~al.
\newblock {From Discovery to the First Month of the Type II Supernova 2023ixf:
  High and Variable Mass Loss in the Final Year before Explosion}.
\newblock {\em ApJL} {\bf 2023}, {\em 955},~L8,
  \href{http://arxiv.org/abs/2307.03165}{{\normalfont
  [arXiv:astro-ph.HE/2307.03165]}}.
\newblock {\url{https://doi.org/10.3847/2041-8213/acf299}}.

\bibitem[{Kozyreva} et~al.(2025){Kozyreva}, {Caputo}, {Baklanov}, {Mironov},
  and {Janka}]{Kozyreva25}
{Kozyreva}, A.; {Caputo}, A.; {Baklanov}, P.; {Mironov}, A.; {Janka}, H.T.
\newblock {SN 2023ixf: An average-energy explosion with circumstellar medium
  and a precursor}.
\newblock {\em A\&A} {\bf 2025}, {\em 694},~A319,
  \href{http://arxiv.org/abs/2410.19939}{{\normalfont
  [arXiv:astro-ph.HE/2410.19939]}}.
\newblock {\url{https://doi.org/10.1051/0004-6361/202452758}}.

\bibitem[{Fuller} and {Tsuna}(2024)]{Fuller24}
{Fuller}, J.; {Tsuna}, D.
\newblock {Boil-off of red supergiants: mass loss and type II-P supernovae}.
\newblock {\em arXiv e-prints} {\bf 2024}, p. arXiv:2405.21049,
  \href{http://arxiv.org/abs/2405.21049}{{\normalfont
  [arXiv:astro-ph.SR/2405.21049]}}.
\newblock {\url{https://doi.org/10.48550/arXiv.2405.21049}}.

\bibitem[{Soker}(2023)]{Soker23}
{Soker}, N.
\newblock {A Pre-explosion Effervescent Zone for the Circumstellar Material in
  SN 2023ixf}.
\newblock {\em Research in Astronomy and Astrophysics} {\bf 2023}, {\em
  23},~081002,  \href{http://arxiv.org/abs/2306.15270}{{\normalfont
  [arXiv:astro-ph.HE/2306.15270]}}.
\newblock {\url{https://doi.org/10.1088/1674-4527/ace51f}}.

\end{thebibliography}

\PublishersNote{}
\end{adjustwidth}
\end{document}